\documentclass{sigchi}
\usepackage{times}
\usepackage{balance}
\usepackage[hyphens]{url}
\usepackage{epsfig}
\usepackage{graphicx}
\usepackage{subfigure}
\usepackage{amsmath}
\usepackage{amssymb} 
\usepackage{amsfonts}
\usepackage{color}
\usepackage{multirow}
\usepackage{algorithm}
\usepackage[utf8]{inputenc}
\usepackage[T1]{fontenc}
\usepackage{cases}
\usepackage{capt-of}

\newenvironment{packed_itemize}{
\begin{itemize}
 \setlength{\itemsep}{1pt}
 \setlength{\parskip}{0pt}
 \setlength{\parsep}{0pt}
 \setlength{\headsep}{0pt}
 \setlength{\topskip}{0pt}
 \setlength{\topmargin}{0pt}
 \setlength{\topsep}{0pt}
 \setlength{\partopsep}{0pt}
}{\end{itemize}}

\newcommand{\para}[1]{{\vspace{3pt} \em \noindent #1 \hspace{10pt}}}
\newcommand{\tabincell}[2]{\begin{tabular}{@{}#1@{}}#2\end{tabular}}

\usepackage{etoolbox}
\makeatletter
\patchcmd{\maketitle}{\@copyrightspace}{}{}{}
\makeatother

\toappear{
Permission to make digital or hard copies of all or part of this
work for personal or classroom use is granted without fee provided that 
copies are not made or distributed for profit or commercial advantage and
that copies bear this notice and the full citation on the first page. To
copy otherwise, or republish, to post on servers or to redistribute to 
lists, requires prior specific permission and/or a fee.\\
{\confname{CHI'12}}, May 5--10, 2012, Austin, Texas, USA.\\
Copyright 2012 ACM 978-1-4503-1015-4/12/05...\$10.00.
}

\makeatletter
\def\url@leostyle{%
  \@ifundefined{selectfont}{\def\UrlFont{\sf}}{\def\UrlFont{\small\ttfamily}}}
\makeatother
\def\pprw{8.5in}
\def\pprh{11in}

\begin{document}

\title{Crowds on Wall Street: Extracting Value from Social Investing Platforms}

% \author{\\ \\ Paper \#155 \\ }

\author{Gang Wang$^\dag$, Tianyi Wang$^\dag$$^\ddag$, Bolun Wang$^\dag$,
Divya Sambasivan$^\dag$, Zengbin Zhang$^\dag$ \\ Haitao Zheng$^\dag$ and Ben Y. Zhao$^\dag$\\
\affaddr{$^\dag$Department of Computer Science, UC Santa Barbara}\\
\affaddr{$^\ddag$Department of Electronic Engineering, Tsinghua University}\\
{\em \{gangw, tianyi, bolunwang, divya\_sambasivan, zengbin, htzheng, ravenben\}@cs.ucsb.edu}}

\newtheorem{theorem}{Theorem}
\newtheorem{definition}{Definition}
\newtheorem{lemma}{Lemma}
\newtheorem{cor}{Corollary}
\newtheorem{fact}{Fact}
\newtheorem{property}{Property}
\newtheorem{remark}{Remark}
\newtheorem{claim}{Claim}

\maketitle
% \thispagestyle{empty}
%%%%%%%%%% add page umber %%%%%%%%%%
\thispagestyle{plain}
\pagestyle{plain}
%%%%%%%%%% add page number %%%%%%%%%%

\begin{abstract}
  For decades, the world of financial advisors has been dominated by large
  investment banks such as Goldman Sachs. In recent years, user-contributed
  investment services such as SeekingAlpha and StockTwits have grown to
  millions of users. In this paper, we seek to understand the quality and
  impact of content on social investment platforms, by empirically analyzing
  complete datasets of SeekingAlpha articles (9 years) and StockTwits
  messages (4 years).  We develop sentiment analysis tools and correlate
  contributed content to the historical performance of relevant stocks. While
  SeekingAlpha articles and StockTwits messages provide minimal correlation
  to stock performance in aggregate, a subset of authors contribute more
  valuable (predictive) content.  We show that these authors can be
  identified via both empirical methods or by user interactions, and
  investments using their analysis significantly outperform broader markets.
  Finally, we conduct a user survey that sheds light on users views of
  SeekingAlpha content and stock manipulation.
\end{abstract}

\section{1. Introduction}
\label{sec:intro}

Social computing tools have touched and revolutionized nearly all aspects of
our daily lives.  Initial impact was focused on person-to-person communication,
where social networks such as Facebook and Twitter displaced emails and
instant messaging.  But 
the impact has spread out far and wide into different aspects of our daily lives,
including job hunting (LinkedIn), blogging (Tumblr), photography (Flickr and
Instagram), travel discovery and rating (Yelp, TripAdvisor).

One domain that has seen dramatic impact is the area of personal investments,
{\em i.e.} buying and selling of stocks, bonds and other investments.  For
decades dating back to the mid-19$^{th}$ century, advice on personal
investments has been the exclusive domain of investment banks and advisors
such as Goldman Sachs, Lehman Brothers and Salomon Brothers.  

Over the last decade, however, the investment landscape has shifted
dramatically in favor of more diversified sources of information.  Networks
like CNBC and Bloomberg established roles as independent sources of financial
news, while the financial crisis of 2008 led to the collapse of several of
the oldest investment banks (Bear Stearns, Lehman Brothers).  Filling in the
void were rapidly growing services such as SeekingAlpha and StockTwits, where
independent analysts and retail investors could contribute and share analysis
for free. SeekingAlpha now reports more than 3 million users and 9 million
unique monthly visits. This potentially represents a significant portion of
the US investment market, with more than 50 million estimated households
that own mutual funds or equities~\cite{ici}. 

In this paper, we seek to understand the quality and impact of opinions and analysis
shared on social investment platforms.  We target the two primary yet quite
different social investment platforms, SeekingAlpha and StockTwits, and
analyze the potential for investment returns following their recommendations
versus the market baseline, the S\&P 500 stock market index.  We seek to
understand how expertise of contributors can affect the quality and utility
of contributed content, using SeekingAlpha as an ``expert'' model (all
content is contributed by less than 0.27\% of users) and StockTwits as a
``peer'' model (any user can contribute).  Our work makes four key
contributions.

{\em First}, we gather longitudinal datasets from both platforms since their
inception (9 years of data for SeekingAlpha, 4 years for StockTwits).  We
develop sentiment analyzers on each dataset, using a mixture of keyword
processing and machine learning classifiers. Validation shows our methods
achieve high accuracy in extracting sentiments towards individual stocks
(85.5\% for SeekingAlpha, 76.2\% for StockTwits).

{\em Second}, we analyze correlation between content sentiment from both
services with stock returns at different time scales.  We show that content
from both SeekingAlpha and StockTwits provide minimal forward correlation
with stock performance.  While the average article provides little value, we
find that a subset of ``top authors'' in SeekingAlpha contribute content that
shows significantly higher correlation with future stock performance.
% find that an investment strategy based on SeekingAlpha sentiment can help
% outperform the market by avoiding some significant crashes, while investments
% guided by StockTwits strictly mirror the S\&P 500 market index.
% guided by SeekingAlpha significantly outperforms the index. The gains come
% from strongly negative sentiments in SeekingAlpha during the crash
% of 2008, sending strong sell signals before the bulk of the market losses.

{\em Third}, we evaluate the hypothetical performance of simple investment
strategies following top authors from both platforms. We show that investment
strategies based on stock sentiment from top SeekingAlpha authors perform
exceptionally well and significantly outperform broader markets. In contrast,
strategies relying on StockTwits generally underperform relative to broader
markets.  In addition, we show that we can identify top authors without
historical stock market data, using only user interactions with their
articles as a guide.

{\em Fourth}, we conduct a large scale survey of SeekingAlpha users and
contributors to understand their usage, reliance, and trust in the
SeekingAlpha service.  Results show that despite seeing potentially
intentionally misleading or manipulative articles, most users still rely
heavily on the site content for investment advice.  Most consider
SeekingAlpha unique, and would not use a competing alternative in its
absence.

A recent article in a financial journal also studied SeekingAlpha and showed
statistical correlation between its content and earning
surprises~\cite{chen2013wisdom}. In contrast, our work contrasts the
performance of expert (SeekingAlpha) versus peer-based (StockTwits) systems,
evaluates the performance of realistic and simple trading strategies, and
reports user views of SeekingAlpha through detailed surveys.

In summary, the rise of crowd-contributed analysis sites has significantly
changed how retail investors manage their investments.  Our analysis shows
that even on curated sites such as SeekingAlpha, broad sentiment is a poor
indicator of market performance.  However, a subset of SeekingAlpha authors
provide valuable content that can be leveraged to build trading strategies
that significantly outperform the broader markets.  More importantly, these
authors can be identified not only by their statistical performance, but more
easily by the feedback their articles generate from other users.  This shows
that even for a complex and domain-specific such as stock trading, broader
input from the crowd can help identify high quality content in a sea of data.
Finally, results from our user survey confirm that most SeekingAlpha users
have seen and have learned to distinguish biased or manipulative articles
from useful articles.

\section{2. Background and Methodology}
\label{sec:back}

% In this section, we briefly describe the background about
% Seeking Alpha and StockTwits. Then we introduce the high-level goals
% of our study. 

%\subsection{2.1 Background: SeekingAlpha and StockTwits }
\para{Seeking Alpha.} Launched in 2004, SeekingAlpha (SA) is the most popular
platform today for independent stock analysis.  As of early 2014, SA has more
than 8 million unique monthly viewers and 3 million registered
users~\cite{alpha1}.  SA's content is mainly contributed by roughly 8000
registered {\em contributors}~\cite{seekingalpha}, and articles are vetted by
an editorial board before publication on the site.  Users can subscribe to
stocks of interest to receive related articles and news summaries, {\em
  follow} contributors to receive their articles, and interact with
contributors and other users through {\em comments} on articles.  SA contributors
include independent investors, portfolio managers, professional investors and
investment firms. Roughly 400 out of 8000 contributors self-identify as
investment firms.  SA pays each contributor \$10 per 1000 page views on
articles.

\para{StockTwits.} StockTwits (ST) started in 2009 as a financial social
network for sharing ideas among traders.  Anyone on StockTwits can contribute
content -- short messages limited to 140 characters that cover ideas on
specific investments, and post their messages to a public stream visible to
all.  There's no editorial board or content curation, and users are not
compensated for their messages.  Like Twitter, ST users {\em follow} others
to build directional social links; and also follow the stock symbols they are
interested in.  Unlike SeekingAlpha, StockTwits provides real-time streaming of
investor sentiment towards individual stocks.  As of the end of 2013,
StockTwits has over 300K registered users and its content reaches an
audience of 40 million across the Internet~\cite{stocktwits}.

\para{Goals and Methodology.} 
Seeking Alpha and StockTwits represent the largest and most representative
sites in expert and peer-based investment analysis. The main goals of this
study are to quantify correlation between sentiment in user-contributed
investment analysis to real movements in stock equities, how and if such
correlations can be leveraged for investment gain, and how users view and
utilize platforms such as SeekingAlpha in their investments.  Our methodology
is as follows:

\begin{packed_itemize}
\item First, we gather complete datasets of contributed articles from
  SeekingAlpha and ``twits'' from StockTwits ($\S3$). We then develop sentiment
  analyzers for both datasets and evaluate their accuracy ($\S4$).
\item Second, we compute statistical correlation between sentiment of
  contributed content to the performance of stocks they discuss ($\S5$). We do so for
  different time scales and both individual stocks and the aggregate
  market.  We also sort authors by their performance to identify authors
  whose content consistently correlate with stock performance.
\item Third, we propose strategies for identifying and trading stocks using
  sentiments from top authors in both platforms, and evaluate them against
  baseline market indices ($\S6$). We explore the efficacy of strategies that
  identify top authors by historical performance and by interactions with
  other users.
\item Finally, we use a large user survey of SeekingAlpha users and
  contributors to understand how they utilize social investing platforms and
  their views on stock manipulation ($\S7$).  Stock ``pump-and-dump'' scams
  have been discovered on SA in the past~\cite{alpha4,alpha3}.
\end{packed_itemize}

\begin{table}[t]
\small{
\centering{
\begin{tabular}{|c|c|c|c|c|c|}
\hline
 	 Site      &
	\tabincell{c}{Data \\Since} &
	\tabincell{c}{Total\\Posts}	&
	\tabincell{c}{Posts w/\\
          Stocks}	&
	\tabincell{c}{Active Users \\ (Authors)}	&
	\tabincell{c}{Covered \\Stocks}	\\
\hline
	SeekingAlpha	& 2004	& 410K	& 163K & 228K (8783)
			& 10.4K	\\
\hline
	StockTwits		& 2009	& 12.7M  & 8.5M	& 86K (86K)	
			& 9.3K	\\
			
\hline
\end{tabular}
\caption{Basic statistics of collected data. }
\label{tab:datasets}
}}
\end{table}

%%% stock diagrams %%%%%%%%
\begin{figure}[t]
 \centering
\begin{minipage}{0.23\textwidth}
\centering
% \vspace{-0.4in}
	\includegraphics[width=0.99\textwidth]{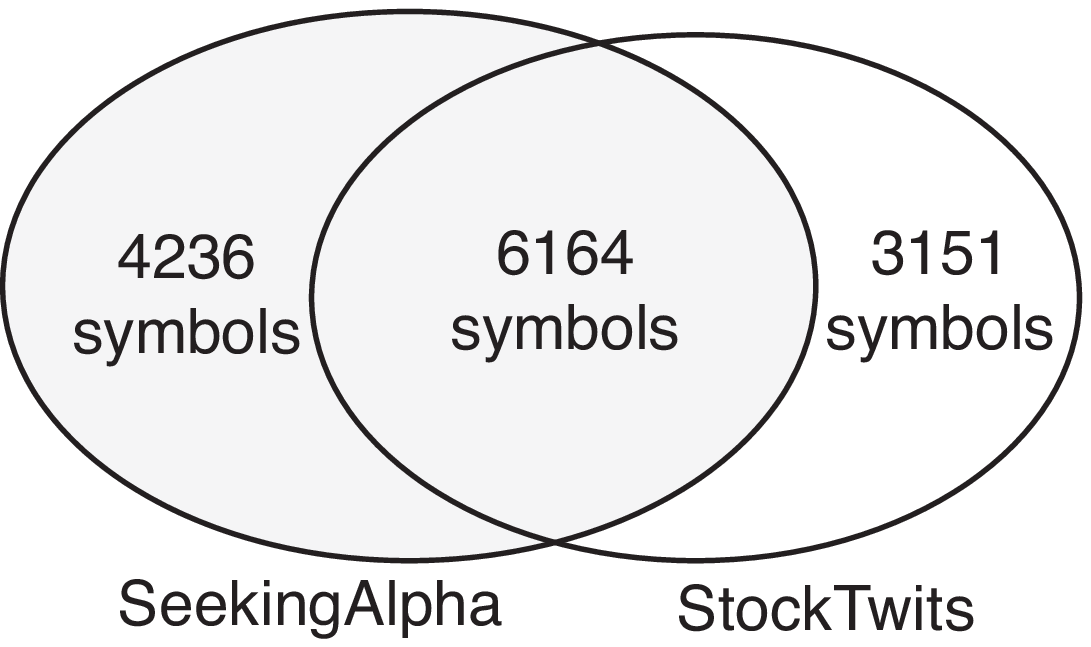}
	%\vspace{-0.45in}
	\vspace{-0.25in}
	\caption{Stock symbols extracted from SA and ST.}
	\label{fig:stock_overlap}
\end{minipage}
\hfill
\begin{minipage}{0.23\textwidth}
\centering
% \vspace{-0.4in}
	\includegraphics[width=0.99\textwidth]{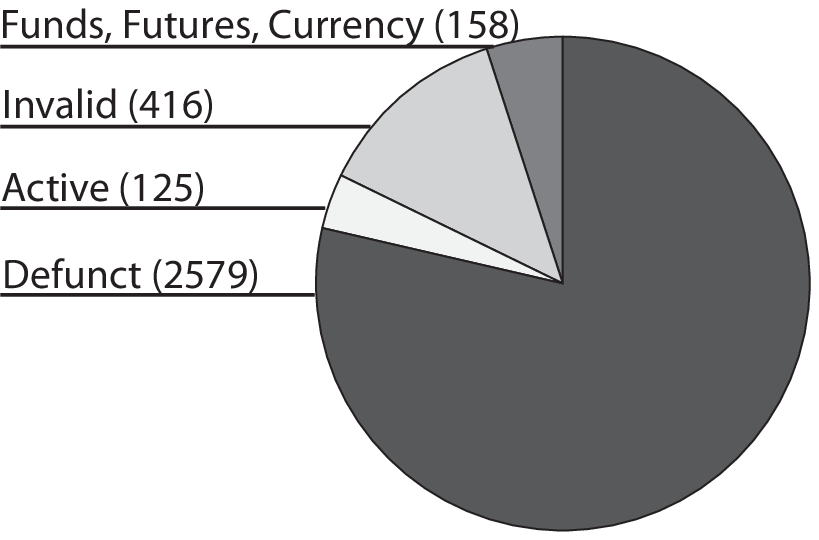}
	%\vspace{-0.45in}
	\vspace{-0.15in}
	\caption{Breakdown of symbols without history price.}
	\label{fig:stock_miss}
\end{minipage}
\end{figure}
%%%%%%%%%%%%%%%%%%%%%%%

\begin{figure*}[t]
\begin{minipage}{0.50\textwidth}
\centering
\subfigure[SeekingAlpha]
{\includegraphics[width=0.48\textwidth]{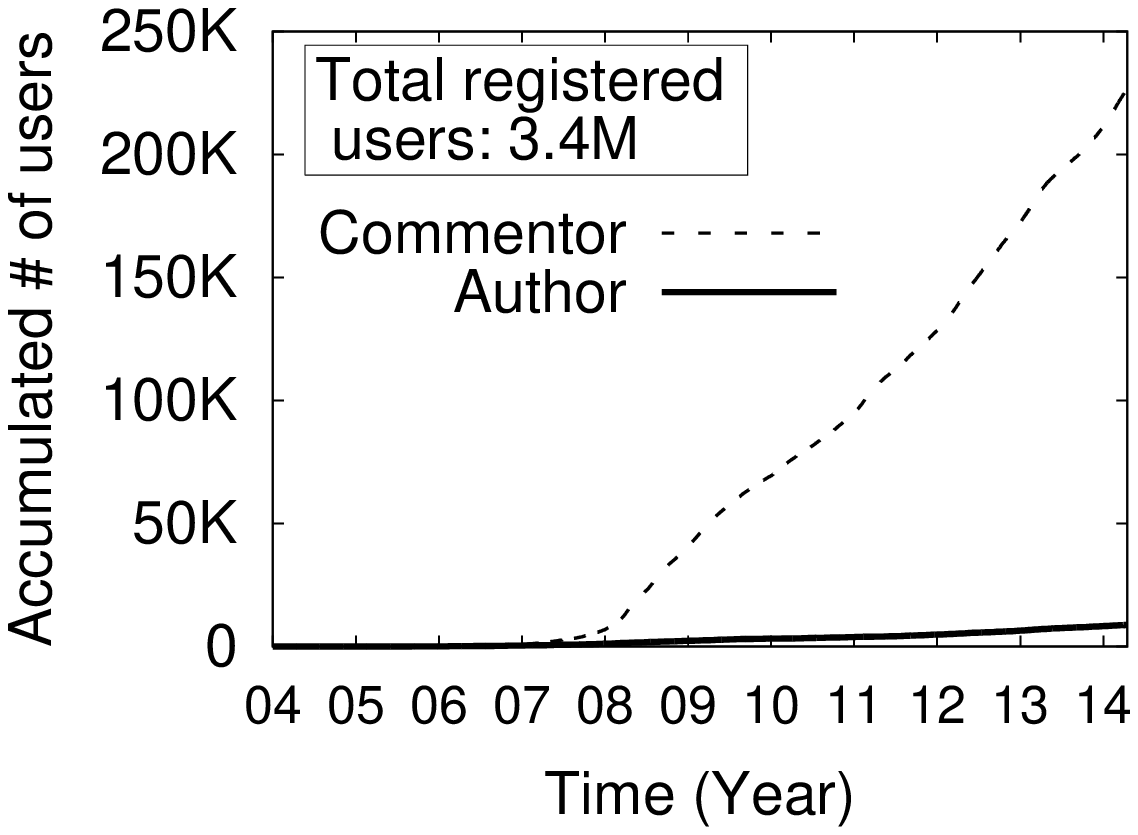}
\label{fig:sa_user_growth}}
\subfigure[StockTwits]
{\includegraphics[width=0.48\textwidth]{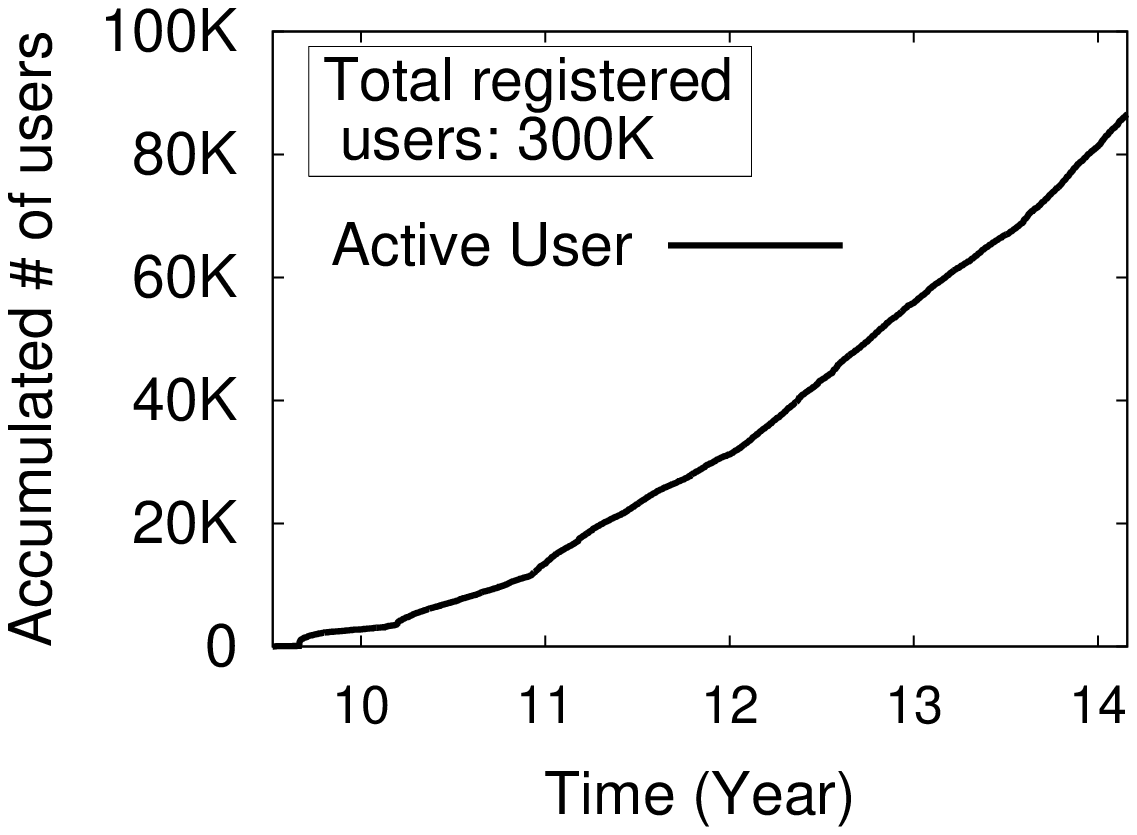}
\label{fig:st_user_growth}}
\caption{Total number of authors and active users over time. The number of
  active users is only a small portion of all registered users.}
\label{fig:user_growth}
\end{minipage}
\hspace{0.02in}
\begin{minipage}{0.50\textwidth}
\centering
\subfigure[Posts by authors.]
{\includegraphics[width=0.48\textwidth]{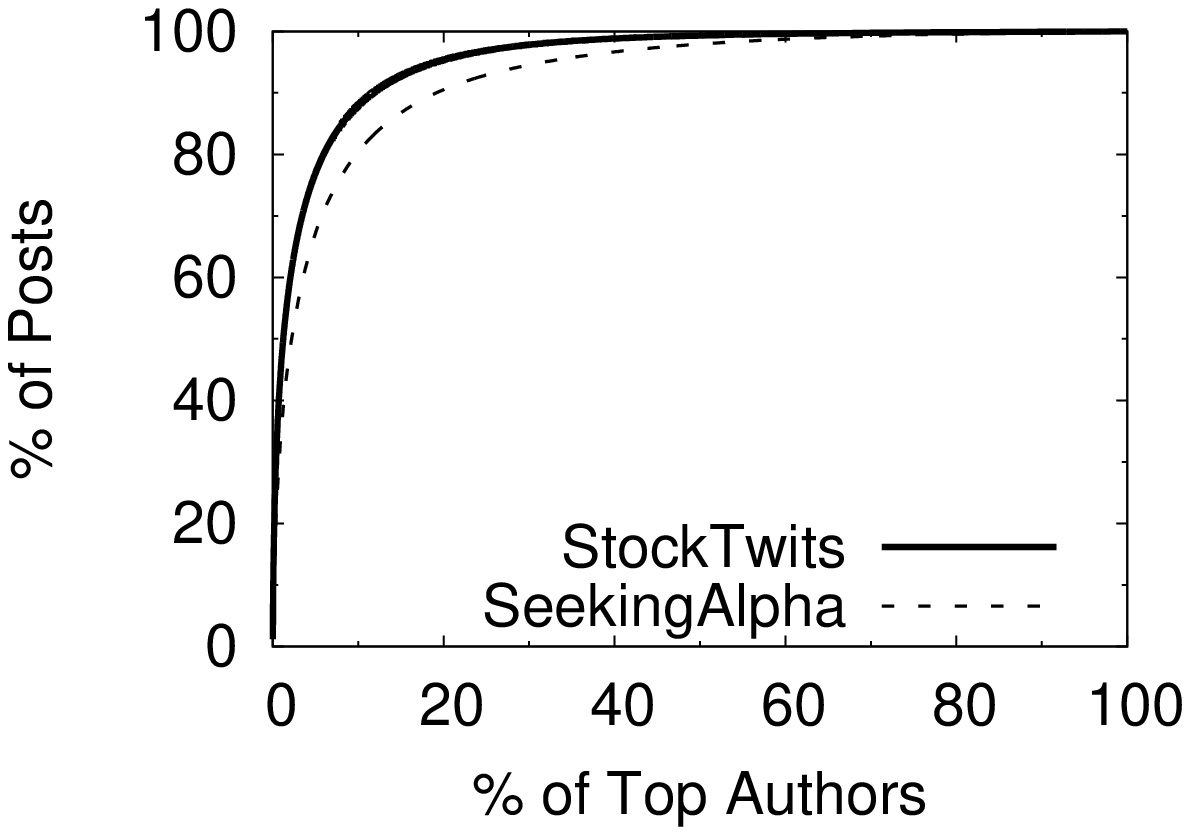}
\label{fig:a_articles}}
\subfigure[Posts per stock.]
{\includegraphics[width=0.48\textwidth]{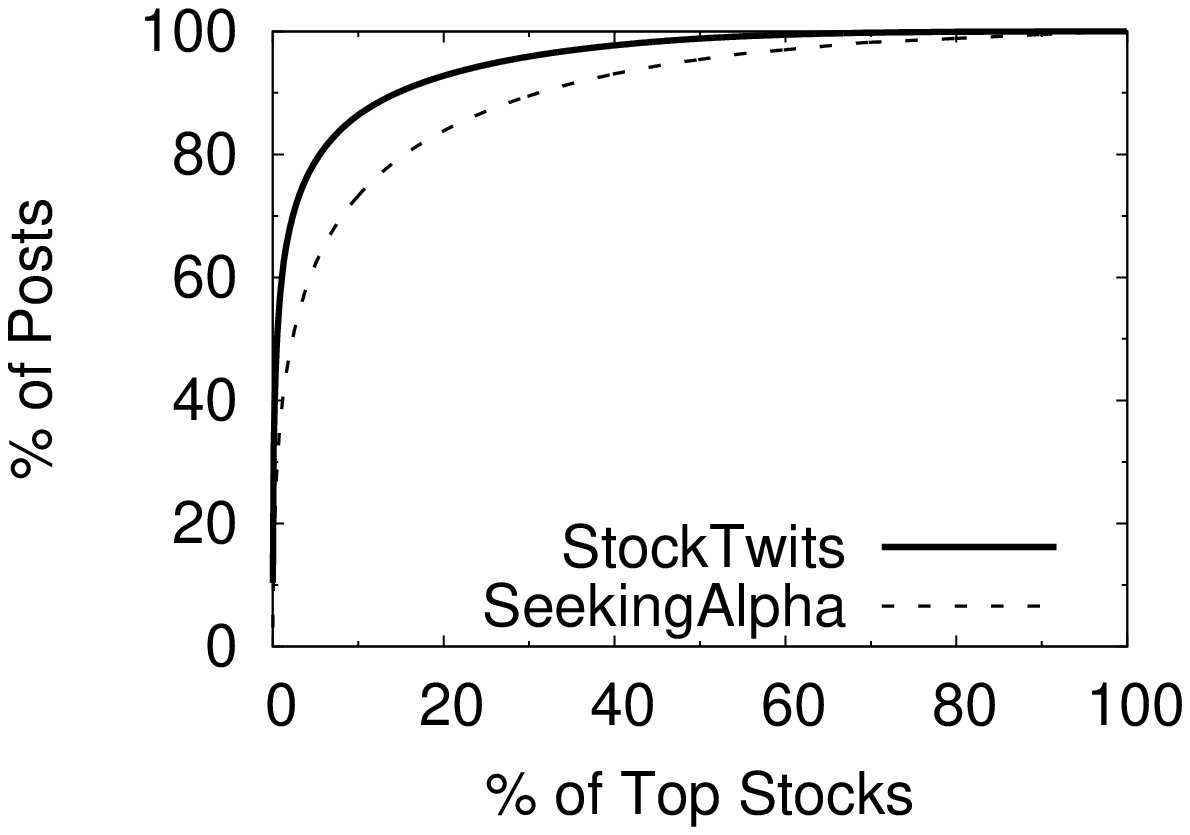}
\label{fig:s_articles}}
\caption{Distribution of articles over authors and individual stocks.}
\end{minipage}
\end{figure*}

\section{3. Data Collection and Initial Analysis}
\label{sec:data}

\subsection{3.1 Data Collection}

\para{SeekingAlpha.} 
In April 2014, we crawled the complete list of historical articles published
on SeekingAlpha since its launch in 2004.  This produced 410,290 articles
written by 8,783 authors, 2,237 news and conference
transcripts\footnote{Transcripts cover board meetings and conference
  calls.}. Our analyses focus on articles and do not include news in the
form of SA ``market currents'' or transcripts.
% editors. In total, our dataset contains 410K articles, from which we extract
% 8,783 unique authors.
Our crawl also produced 4,115,719 total comments, of which 75\% were written by
227,641 non-contributing users. The remaining 25\% are from authors
themselves. We crawl profiles of all authors and active users for
their brief bio and number of followers and followees.

Each SeekingAlpha article has an ``about'' field that lists what stock(s) the
article discusses. 163,410 (about 40\%) of our articles have at least one
stock symbol in their about field. Articles without stock symbols usually
discuss overall market trends or sectors of stocks.  From our entire dataset of
SeekingAlpha articles, we were able to extract 10,400 unique stock symbols.

\para{StockTwits.} We are fortunate to receive permission from 
StockTwits Inc. to access their historical message archive, including 
all messages posted from 2009 (initial launch) to February 2014.  The 
dataset contains 12,740,423 messages posted by 86,497 users. Each 
message includes a messageID, author's
userID, author's number of followers (followees), timestamp, and message
text. Each message is limited to 140 characters, and stock symbols
are preceded by a ``CashTag'' (\$).  In our dataset, about 67\% of
StockTwits messages have at least one CashTag.
% (9.5\%), we again attribute it to all the stocks it talked about.
From these messages, we extract 9,315 unique stock symbols.

StockTwits messages can also be labeled ``bullish'' or ``bearish'' by the
author to label their sentiment towards the mentioned stocks.  10\% 
of messages (1.3 million) have this label.  We use these labeled messages later as 
ground-truth to build and evaluate sentiment analysis tools.

\para{Stock Historical Price Data. } 
Our two datasets include a total of 13,551 {\em unique} stock symbols.
Symbols from two sites do not completely overlap
(Figure~\ref{fig:stock_overlap}): 6,164 symbols appear in both datasets, most
represent stocks on the NASDAQ and NYSE exchanges.  SeekingAlpha-only symbols
(4,236) are mostly small stocks sold on Over-The-Counter Bulletin Board
(OTCBB), while StockTwits-only symbols (3,151) are mostly from the Toronto
Stock Exchange.

We use the Yahoo! Finance open API~\cite{yahoo_finance_api} to crawl
historical prices for all stock symbols. For each stock, we obtain its
historical {\em daily} opening and closing prices, volume, and intraday price
range.  Of our 13,551 symbols, we found data for 10273 symbols.  We track
down the 3278 missing symbols using both Yahoo Finance and Bloomberg (see
Figure~\ref{fig:stock_miss}). First, 2579 missing symbols are defunct, {\em i.e.}
symbols made invalid due to corporate breakups, merge/acquisitions, or
bankruptcy. Second, 125 are active stocks, either on foreign exchanges or OTC
stocks not covered by Yahoo. Third, 158 symbols are ETF or mutual Funds,
Futures contracts, and Currencies. Finally, we manually inspect the remaining
416 symbols, and find they are often user-defined symbols such as {\tt
  \$CRASH}, or non-listed companies such as {\tt \$QUORA}. Missing symbols
account for 7\% of SeekingAlpha articles and 6\% of
StockTwits messages, thus we believe it would not impact our overall
conclusion. We summarize our final datasets used in our analysis in
Table~\ref{tab:datasets}.

\begin{figure}[t]
 \centering
\begin{minipage}{0.48\textwidth}
 \centering
	\includegraphics[width=1\textwidth]{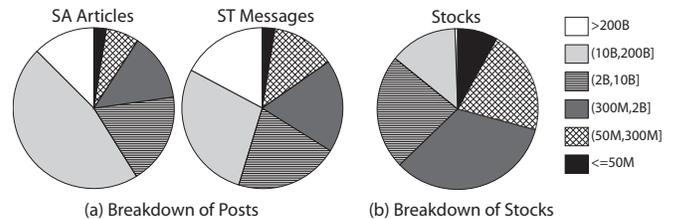}
%	\vspace{-0.1in}
	\caption{\% of posts/stocks in different market cap. categories.}
	\label{fig:cap_count}
\end{minipage}
% \vspace{-0.03in}
\end{figure}

\subsection{3.2 Preliminary Analysis}
\label{preliminary_analysis}
Here, we briefly analyze our datasets to understand the structure of user
communities and content in these two systems.  The two systems are quite
different. SeekingAlpha focuses on detailed, blog-like contributions by a
small group of ``experts'' further curated by editors, and the large majority
of users consume and comment on articles. In contrast, StockTwits encourages
short, terse contributions by all users.  We compare and contrast the
platforms on growth over time, skew of author contributions, distribution of
stocks covered, and the structure of social connections in the community.

\para{User Growth.}  Figure~\ref{fig:user_growth} plots the
growth of users over time for both systems. Recall our SeekingAlpha data
includes all users who have contributed or commented on at least 1 article.
At both sites, active users are growing at a stable rate, but only make up a
small portion of all registered accounts (236,000 active versus 3.4 million
accounts in SeekingAlpha, and 86,000 active versus 300K accounts for
StockTwits).

\para{Distribution of content over authors and stocks.} Author's contribution
to the platform is measured by the number of articles from the author. For
both SeekingAlpha and StockTwits, we find the contribution per authors is
highly skewed (Figure~\ref{fig:a_articles}). On SeekingAlpha, 20\% of the
most active users contributed 80\% of articles, while on StockTwits, 20\% of
active users contributed 90\% of the messages. Even though StockTwits tries
to leverage the power of the crowd, it has an even higher skewness in content
contribution than SeekingAlpha. This sheds concerns that the wisdom of the
crowd is likely to be dominated by the most active authors.

Content posted on both sites is also highly skewed to a small
portion of ``popular'' stocks (Figure~\ref{fig:s_articles}). More than 70\%
of SeekingAlpha articles cover top 10\% of the most popular
stocks. The skew is even stronger in StockTwits, with 90\% of 
messages focusing on 10\% most popular stocks.

Figure~\ref{fig:cap_count} shows the heavy emphasis of articles on large
capitalization companies.  47\% of SeekingAlpha articles and 28\% of
StockTwits messages cover stocks of companies between \$10 Billion and \$200
Billion in market cap, which account for only 14\% of all stocks. The
emphasis is stronger for the largest companies (market cap $>$\$200
Billion).  They account for only 0.3\% of all stocks, but are covered by
10-15\% of the content on both platforms.

\begin{table}[t]
  \centering
  \small{
\begin{tabular}{|c|c|c|c|c|c|c|c|}
  \hline
  Graph & Nodes & Edges & \tabincell{c}{Avg. \\ Degree} & \tabincell{c}{Cluster.\\Coef.} & \tabincell{c}{Avg.\\Path} & Assort. \\

  \hline
  SeekingAlpha & 386K & 3.9M & 19.91 & 0.150 & 3.09 & -0.428 \\
  \hline
  Facebook~\cite{interaction} & 1.22M & 121M & 199.6 & 0.175 & 5.13 & 0.17 \\
  \hline
  Orkut~\cite{socialnets-measurement} & 3.07M & 224M & 145.5 & 0.171 & 4.25 & 0.072 \\
  \hline
  Flickr~\cite{socialnets-measurement} & 1.85M & 22.6M & 24.5 & 0.313 & 5.67 & 0.202 \\
  \hline
\end{tabular}
}
\caption{Comparing the structure of social graphs from SeekingAlpha and more
  traditional social networks.}
\label{tab:graph}
\vspace{-0.1in}
\end{table}

\para{Social Connections and Graphs.} We analyze social connections in
SeekingAlpha to understand the structure of the network (similar
data was not available for StockTwits). We crawled a full snapshot of the
SeekingAlpha network in October 2013, using contributors as seeds
and crawling all reachable users. The result graph has 386K nodes and
3.9M edges.

We compute key graph metrics and compare them against those of popular social
networks in Table~\ref{tab:graph}.  The Facebook graph is from the Australia
regional network in the reference paper~\cite{interaction}. First,
SeekingAlpha has a similar clustering coefficient to other social networks,
indicating that SeekingAlpha has a similar level of local connectivity.
Second, SeekingAlpha has a dramatically different (negative) assortativity
value compared to other networks. Assortativity measures how strongly users
tend to connect to others with similar degree of connectivity.  Positive
assortativity indicates a preference to link to users with similar degree,
while negative assortativity indicates a preference to link to users with
different degree.  SeekingAlpha's extremely negative assortativity (-0.428)
indicates that SeekingAlpha exhibits a highly bi-partite structure, where
most connections are between users following contributors, but with fewer
connections amongst accounts of the same type.  Finally, the very low average
path length in SeekingAlpha is indicative of networks where supernodes
produce short connections between users.

\section{4. Sentiment Extraction}
\label{sec:tool}
Our analysis on the value of user-contributed investment analysis hinges on
our interpretation of sentiment in SeekingAlpha articles and StockTwits messages.
The first step in this process is developing reliable tools to extract
sentiment (positive or negative opinion) on stocks from posted articles and
messages.  We discuss our sentiment analysis techniques here, and rely on
them in later sections to compute stock performance correlation and to drive
trading strategies.

Our approaches to extract sentiment from SeekingAlpha and StockTwits are
quite different. More specifically, SeekingAlpha articles are sufficiently
long to apply an approach using a keyword dictionary, while we applied a
supervised machine learning approach for the short messages in StockTwits,
using messages with ``bullish'' or ``bearish'' labels as training data.  Our
validation results show we achieve an accuracy of 85.5\% for SeekingAlpha
and 76.2\% for StockTwits.  We note that these accuracy results are on par or
significantly better than existing sentiment analysis
techniques~\cite{chen2013wisdom,gonccalves2013comparing}.

\begin{figure*}[t]
 \centering
 \subfigure[SeekingAlpha (2005-2014)]{
  \label{fig:return_pear1}
  \includegraphics[width=0.4\textwidth]{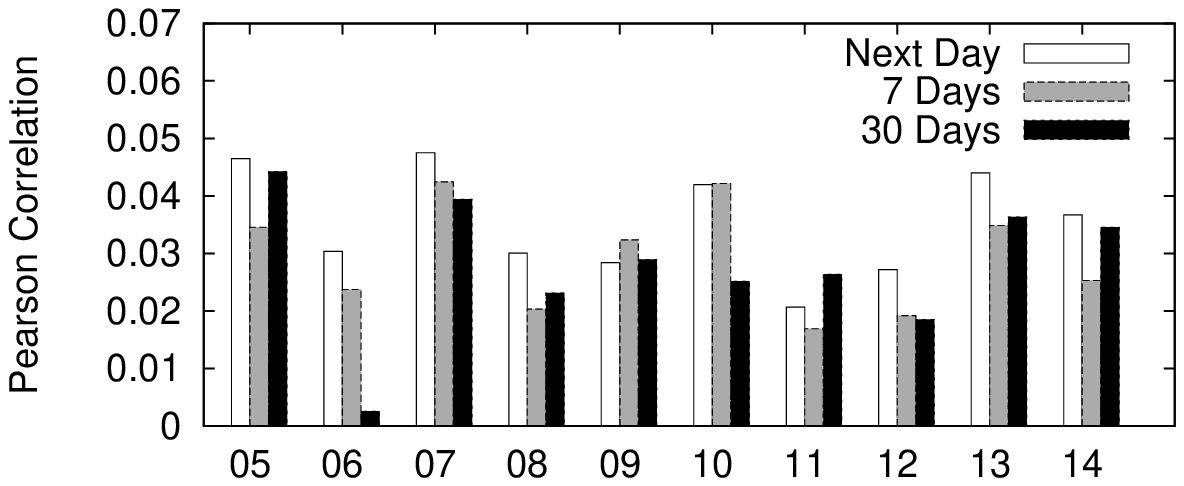}}
\hspace{0.2in}
  \subfigure[StockTwits (2009-2014)]{
  \label{fig:return_pear2}
  \includegraphics[width=0.4\textwidth]{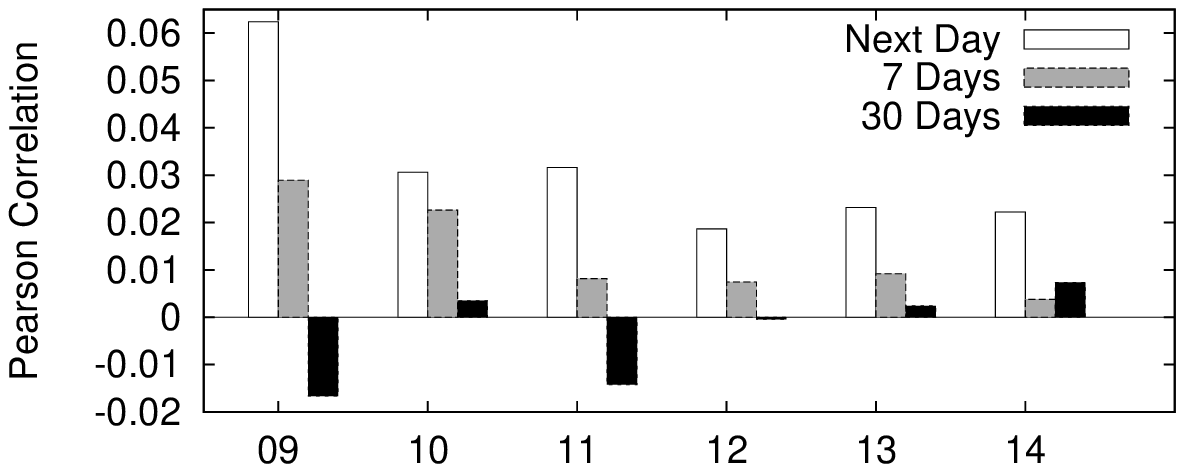}}
\vspace{-0.01in}
\caption{Pearson correlation coefficient between article sentiment and
          near term stock price movement (in 1, 7, 30 days).}
\label{fig:return_pear}
\vspace{-0.03in}
\end{figure*}

\subsection{4.1 Sentiment Analysis: Seeking Alpha}
We develop a dictionary based method to extract sentiment from
SeekingAlpha articles. At a high level, we measure author sentiment (towards a
stock) based on the ratio of positive and negative keywords in the
article.  We rely on a widely used financial sentiment
dictionary~\cite{loughran2011liability} to identify
positive and negative keywords in the article, and calculate the
sentiment score as $S=log \frac{1+\sum P_{i}}{1+\sum N_{i}}$, where $P_{i}$ ($N_{i}$) is the
number of positive (negative) words or phrases in sentence $i$.  
The sentiment score is a decimal value with high positive value indicating
strong and positive sentiment, and vice versa. For example,
an extremely positive article with 100 positive words and 0 negative
words gets a score of 4.6; an extremely negative article with 100
negative words and 0 positive words scores -4.6.

However, there are problems with applying this method naively to SeekingAlpha
articles. First, many articles discuss multiple stocks, and sentiments may be
quite different for each discussed stock. Generating one sentiment score for
all stocks is clearly oversimplifying.  Second, simple
keyword counting can easily make mistakes. For instance, ``low risk''
contains negative keyword ``risk'' but the overall sentiment is
positive. Also negation usually changes the sentiment of words, such as ``not
good,'' or ``no benefits.''

We make several refinements to our method to address these challenges. 
First, we partition multi-stock articles and assign individual sentences to each
stock symbol\footnote{Stock symbols are easily recognized as hyperlinks
   in each article.}. Our method is a simple
distance-based slicing: we consider stock symbols (and its company
names) as landmarks, and we assign each sentence to the closest
landmark in the article. Next, we make two adjustments to basic
keyword counts. First, we identify the sentiment of noun
phrases such as ``higher revenue,'' ``low return,'' ``low risk.'' We
extract frequent noun phrases that occur in more than
1\% of articles, and manually label their sentiment. Second, we reverse
the sentiment of words or phrases affected by negation
words~\cite{pang2002thumbs}.

To validate our method, we sample 300 articles from our article collection
and manually label their sentiment as positive or negative. We have three
graduate students read each article and vote for the final label. Then we run
our sentiment extraction method on these articles to generate sentiment
scores. The result shows our method achieves 85.5\% accuracy.  Note that this
accuracy only considers the polarity of the scores, {\em i.e.} whether an
article is positive or negative.

\subsection{4.2 Sentiment Analysis: StockTwits}
Roughly 10\% of StockTwits messages already have sentiment labels, either
``bullish'' or ``bearish.''  Our goal is to extract sentiment for the
remaining 90\% of messages. We choose to use supervised Machine Learning
method, since messages are too short for dictionary-based approaches
(we confirmed this with experiments). 

To build a machine learning classifier, we follow prior
work~\cite{pang2008opinion} to use the existence of unigrams as features. To
reduce noise in the machine learning model, we exclude infrequent unigrams
that occur less than 300 times over all messages, and remove stopwords, stock
symbols and company names from messages.  We use the ground-truth messages as
training data, and empirically test multiple machine learning models,
including Naive Bayes, Supported Vector machine (SVM) and Decision
Trees.  We randomly sample 50K messages labeled as
``bearish'' and 50K labeled as ``bullish,'' and run 10-fold cross
validation.  We find the SVM model produces the highest accuracy (76.2\%), and
use SVM to build the final classifier used on all messages.

The sentiment score of StockTwits messages is binary: 1 indicates positive sentiment and -1 indicates
negative sentiment.  For rare messages with multiple symbols, we
attribute the same sentiment score to all symbols in the message
(messages are too short to slice).

\section{5. Predicting Stock Price Changes}
\label{sec:model}

Using our sentiment analysis tools, we can now quantify the value of
SeekingAlpha and StockTwits content, by measuring statistical correlation
between their sentiment towards individual stocks, and each stock's
near-term price performance.  Our goal is to study such correlation for
different time periods after the content was published, for both platforms
and over different historical periods (to account for bull/bear cycles in the
stock market).

% changes. This analysis is to gauge the overall usefulness of
% user-generated sentiment and lay the ground work for developing
% investment strategies in later sections. In this section, we
% quantitatively exame the correlation between the sentiment of the crowd
% and the future trend of stock price. More specifically, we want to
% understand whether such correlation exists, how strong it is across
% different market years, and what's the rough
% time window for the sentiment to be effective. With these questions in
% mind, we measure the correlation at both the {\em per-article} level
% as well as the network-wise {\em aggregated} level. 

% \subsection{5.1 Article Sentiment vs. Individual Stock}
\subsection{5.1 Per-article Sentiment and Stock Performance}
We start by studying how well each article predicts the future price trends
of the stocks it discusses.  We compute the Pearson correlation
coefficient~\cite{Pearson1895ab} between article's sentiment (positive or
negative) and stock's future price change.  For simplicity, we ignore
magnitude of price movements and strength of sentiments, and reduce both
metrics to binary (positive/negative) values.  Pearson correlation
coefficient is widely used to measure the linear correlation between two
variables\footnote{ For two variables $X$ and $Y$, the Pearson correlation
  coefficient $ \rho _{X,Y} = \frac{E[(X-\mu_X )(Y-\mu_Y)]}{\sigma_X \sigma_Y
  } $, where $\mu_X$ and $\sigma_X$ is the mean and standard deviation of $X$
  and $E$ is the expectation.}. Its value ranges from -1 to 1, where 1 means
perfect positive correlation, 0 means no correlation, and -1 means perfect
negative correlation. In this context, the Pearson coefficient is 1 if a
stock always increases in value after it is discussed positively by an article.

We compute the Pearson coefficient between two variables $S$ and $P$. $S$ is
1 (-1) if the article's sentiment is positive (negative); and $P$ is 1 or -1
depending on whether the discussed stock goes higher or lower in price.  We
study stock price changes in different time windows after a relevant article
is published, including the next day, the next week and the next month.  For
articles with multiple stock symbols, we count each stock as one data point.
We also group articles in each year and compute the per-year Pearson
correlation coefficient to understand the consistency of correlation across
different years.  Results are shown in Figure~\ref{fig:return_pear}.

% To better interpret the Pearson coefficient value, we set up some ground-truth
% baselines. For example, if the articles predict the future stock trend
% with an accuracy of 75\% (assuming equal accuracy for predicting
% price going up and down), the corresponding correlation coefficient is
% about 0.4; Similarly, an accuracy of 65\% maps to a correlation score 0.15, and 53\%
% maps to 0.05. {\em i.e.} positive
% (negative) sentiment predicts price going up (down), 

First, we observe that correlation is extremely low across different time
windows and different market years for both systems. To better understand the
Pearson values, consider that a prediction history of 75\% correlation would
produce a Pearson coefficient of 0.4.  The most significant correlation in
our results is 0.05, which translates to a prediction accuracy of 53\%, 3\%
better than a random guess. This means that taken as an aggregate,
SeekingAlpha articles and StockTwits messages provide minimal value for
investors.

Looking closer, SeekingAlpha generally does a bit better than random, while
StockTwits has weaker, sometimes negative correlations. Clearly StockTwits is
better as a gauge of instantaneous market sentiment, and a poor predictor of
even near-term performance.  In contrast, SeekingAlpha is a bit more
consistent over different time windows, and still has some value for
predicting price movements in the following month.  We note that SeekingAlpha
accuracy is a bit lower for market years with high volatility (2008--2009,
2011--2012).

\begin{figure}[t]
\centering
  \subfigure[Top authors ranked by 2013.]{
  \label{fig:topauthor1}
  \includegraphics[width=0.23\textwidth]{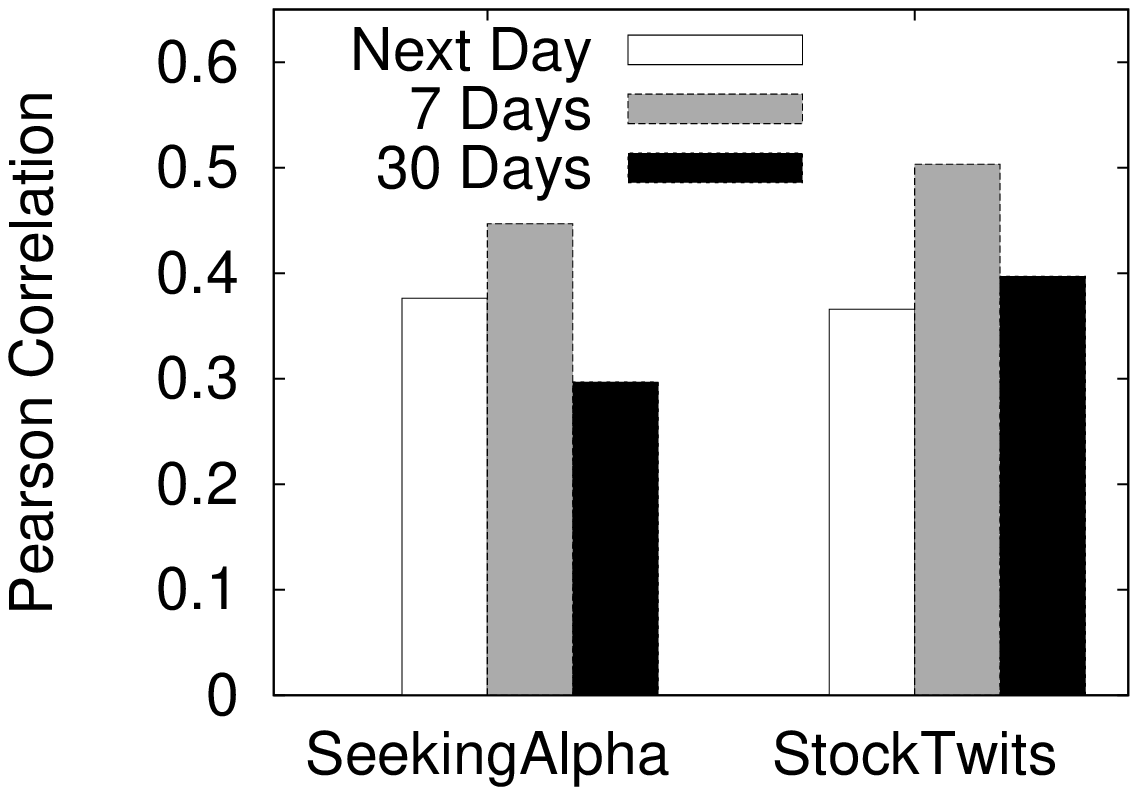}}
  \subfigure[Top authors ranked by 2012.]{
  \label{fig:topauthor2}
  \includegraphics[width=0.23\textwidth]{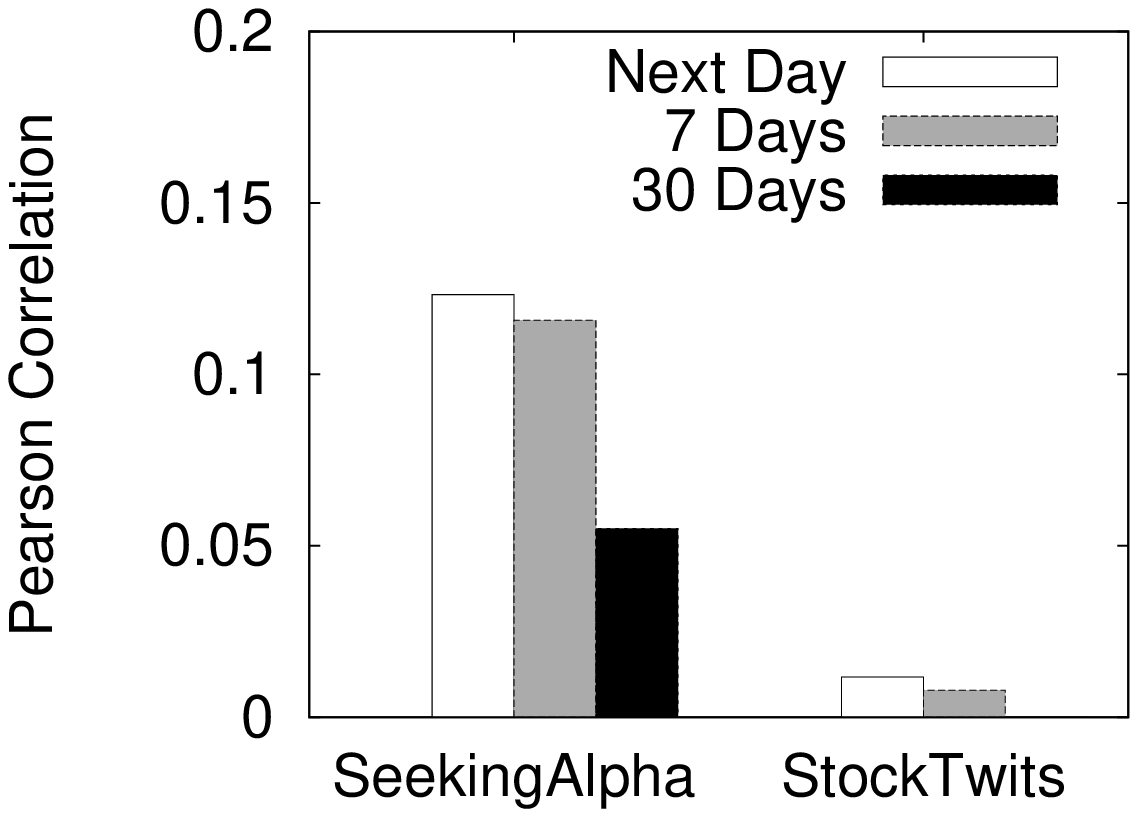}}
\caption{Correlation analysis of stock performance in 2013 by top authors. We
  consider two sets of authors, top authors based on performance in 2013
  (left) and based on performance in 2012 (right). }
 \label{fig:topauthor}
\vspace{-0.15in}
\end{figure}
\begin{figure*}[t]
\begin{minipage}{0.64\textwidth}
 \centering
 \subfigure[SeekingAlpha (2005-2014)]{
  \label{fig:return1}
  \includegraphics[width=0.48\textwidth]{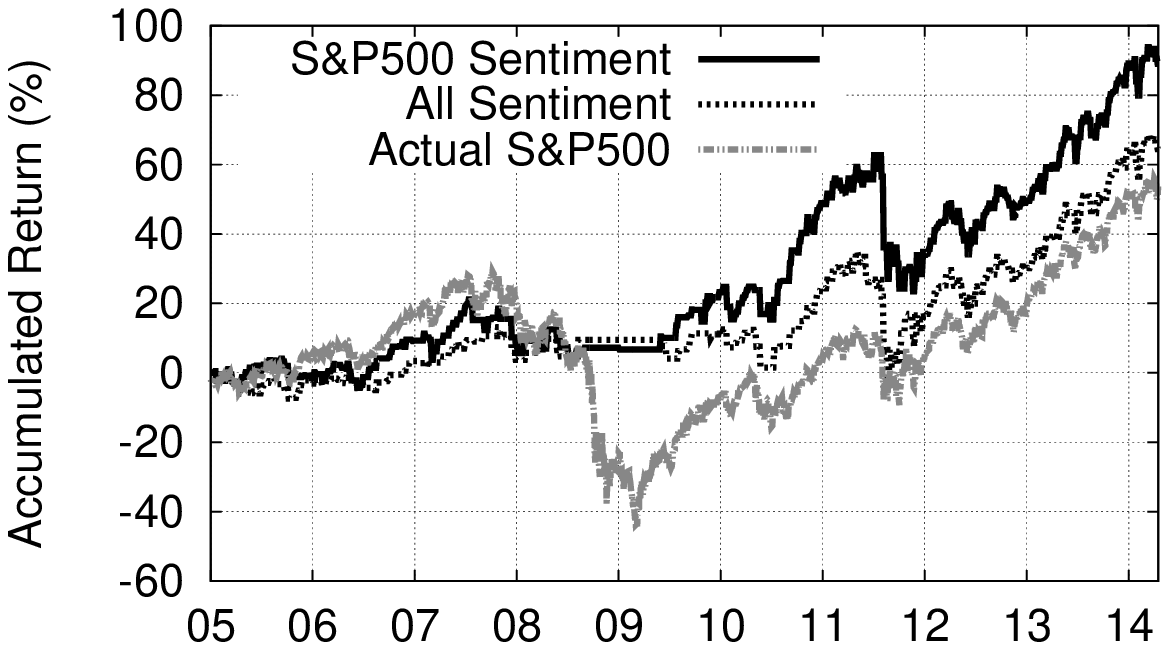}}
\vspace{-0.01in}
  \subfigure[StockTwits (2009-2014)]{
  \label{fig:return2}
  \includegraphics[width=0.48\textwidth]{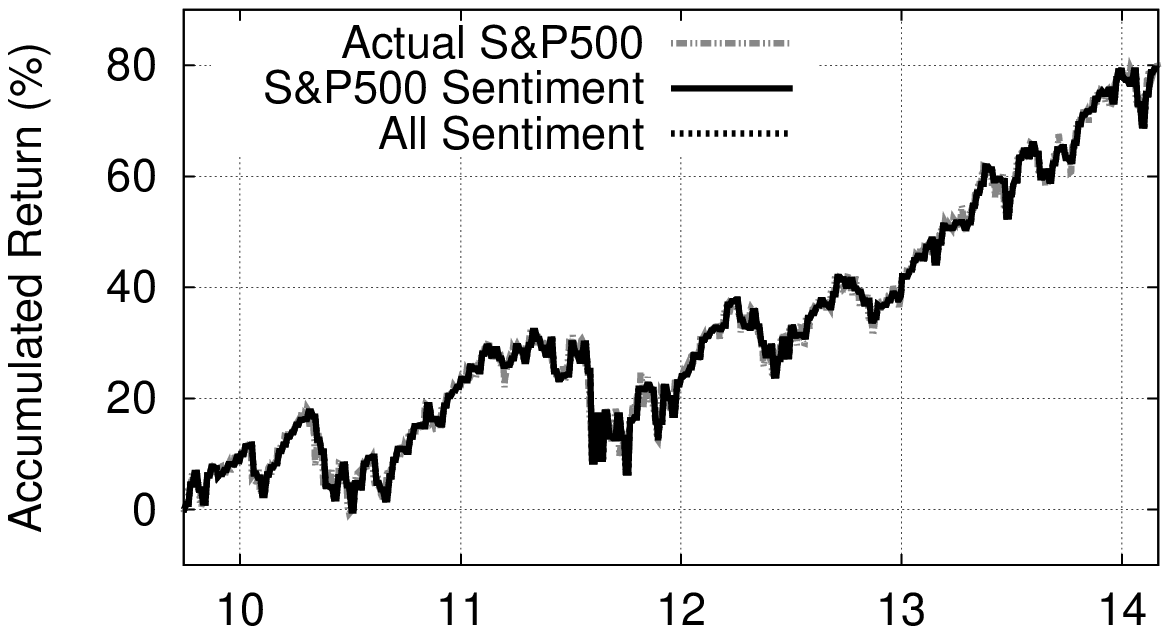}}
\vspace{-0.01in}
\caption{Predicting S\&P 500 index using aggregated sentiment of whole
  site versus only using sentiment related to the indexed 500 stocks.}
\label{fig:return}
\end{minipage}
\vspace{-0.03in}
\hfill
\begin{minipage}{0.32\textwidth}
 \centering
	\includegraphics[width=1\textwidth]{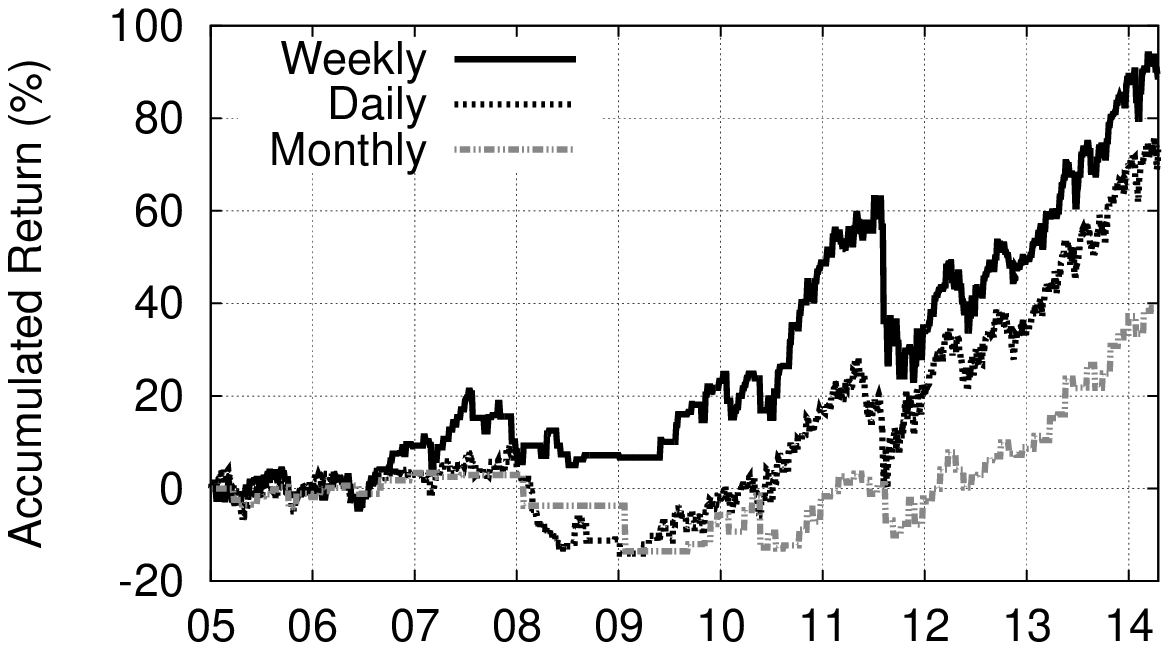}
	\vspace{-0.15in}
	\caption{Impact of trading time window (SA).}
	\label{fig:tw1}
\end{minipage}
\vspace{-0.15in}
\end{figure*}

The natural followup question is: is the weak correlation consistent across
all authors, or are there authors whose contributions provide consistently good
predictors of stock performance but are buried in the noise?  Here, we
perform a simple test to determine if more accurate authors exist.  We
rank authors based on how well their articles predict stock
returns in a single year, {\em e.g.} 2013. For each author, we measure the
average hypothetical return per article for all her articles as a percentage
after a time window $W$. If $P(x)$ is the stock closing price of a given day $x$,
and the article is posted on day $d$, then return $R$ from a positive article
is $R = \frac{P(d+W)-P(d)}{P(d)}$, and return on a negative article
is $R = -\frac{P(d+W)-P(d)}{P(d)}$. 

For our simple experiment, we set $W$ to one week, and compute the average
return per article to rank authors in 2013.  We then take a closer look at
correlations of 500 stocks discussed by the top ranking authors.
Figure~\ref{fig:topauthor1} clearly shows that correlation scores for top
authors in SeekingAlpha and StockTwits are both very high (around 0.4), {\em
  i.e.} top authors can predict stock movement within a week with $\sim 75\%$
accuracy.  
% This result shows that there's a potential to recover useful sentiment by
% selecting reliable authors and removing noises in the crowd.

In reality, we cannot identify top authors using data from future months in
the same year.  We can only rely on past performance to guide us.  Thus we
repeat the experiment: using 2012 data to rank authors, and then we study the
performance of their 2013 stock recommendations.  As expected, correlation
results for those authors' stocks in 2013 (see Figure~\ref{fig:topauthor2})
are much lower. Top SeekingAlpha authors show a significant correlation score
around 0.12, which is still significantly better than the average.  This
confirms our intuition, that filtering out the ``noise'' does indeed reveal
more accurate contributors in the crowd.  Note that the same does not hold
for StockTwits, {\em i.e.} no StockTwits authors can consistently predict
stock performance over different time periods.
% We will discuss more regarding to how to leverage top authors to develop
% investment strategies in later section.

% We now intend to examine whether the power of the crowd (aggregating
% sentiment of the whole network) can help to compensate errors of
% individual articles and produce more accurate predictions.
% \subsection{5.2 Aggregated Sentiment vs. Market Trend} 
\subsection{5.2 Aggregated Sentiment for Market Prediction} 
Since the correlation between articles and individual stocks is weak, we
consider aggregated sentiment of all articles as a possible predictor for the
market as a whole.  Here we use the S\&P 500 index\footnote{S\&P 500 is a
  widely-accepted market index based on the market capitalizations of 500
  large companies.} as a metric for overall market performance.  Here, we
treat the S\&P 500 index as a single stock and trade it based on the {\em
  aggregated} sentiment over time. In practice, this can be done using {\tt
  SPY}, an Exchange Traded Fund (ETF) that tracks the S\&P~500. 

The process is intuitive: we start by holding a position in the S\&P
500.  After every $K$ days, we check the aggregated sentiment of articles posted
in the past $K$ days, and choose to either buy or sell the S\&P~500. 
Sentiment over a window of $K$ days is computed by first computing the
average sentiment of all articles for each day, then setting the overall
sentiment to positive if there are more net-positive days than net-negative
days. If the sentiment for a window is negative, we sell our entire position (if
any) in the S\&P~500.  If the sentiment is positive, we buy back our full
position if we have none, or hold it if we already have a position.
% Then our return between trading day $t$ and next trading day $t+K$ is
% $R(t) = \log(\frac{P(t+K)}{P(t)})$ if we hold the stock, where
% $P(t)$ is the index closing price on day $t$; Automatically, the
% return is 0 if we don't hold the stock. 
% The trading will make profit if the sentiment in
% previous $K$ days can correctly predict the trend of S\&P 500 index in
% the next $K$ days. We will test different $K$ later to see its impact. 

\para{Long-term Performance.} We simulate this trading strategy using data
from SeekingAlpha (January 2005 to March 2014) and StockTwits (September 2009
to February 2014)\footnote{SeekingAlpha only had 3 total articles in
  2004. Thus we start our SeekingAlpha simulations from 2005.}.  We set time
window $K$ to {\em one week} (we evaluate $K$'s impact later). On each
dataset, we run two configurations, one using ``all'' sentiment from the
entire network, and the other only taking sentiment specifically about the 500
stocks listed in S\&P 500 index\footnote{The stocklist of S\&P 500 index
  changes periodically, and we adapt the list in our evaluation
  accordingly.}. As a baseline, we run a ``buy-and-hold'' strategy on S\&P
500 index, that is, holding the stock for the entire duration of the
simulation.

Figure~\ref{fig:return} plots the total return (normalized by initial
investment) accumulated over time. For SeekingAlpha
(Figure~\ref{fig:return1}), we find both configurations outperform the actual
S\&P~500.  Aggregated sentiment can generally predict market trends.  Not
surprisingly, sentiment specifically about the 500 stocks in the index
produces more accurate results.
%  Second, we find SeekingAlpha sentiment surpasses the baseline
% by $>$ 30\% of return. 
A closer look shows that our strategy significantly outperforms the real
market during 2008--2010, when the financial crisis caused the stock market
(and the S\&P 500) to lose more than half of its value. Given the overall
negative sentiment in SeekingAlpha, our strategy held no positions and
avoided much of the market losses.
For StockTwits (Figure~\ref{fig:return2}), we find that all three lines
completely overlap. In fact, after we aggregate the sentiment of the whole
network, StockTwits's overall opinion towards the market is almost always
positive.  Our sentiment-driven trading is equivalent to
buy-and-hold. 

\begin{figure}[t]
\centering
\begin{minipage}{0.32\textwidth}
	\includegraphics[width=1\textwidth]{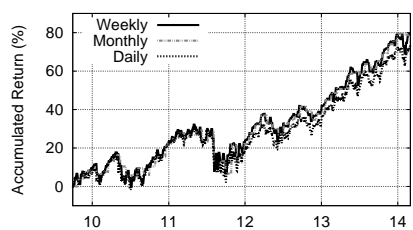}
	\vspace{-0.1in}
	\caption{Impact of trading time window (ST).}
	\label{fig:tw2}
\end{minipage}
\vspace{-0.03in}
\end{figure}

\para{Impact of Time Window $K$.} To understand the time frame $K$ for
aggregating sentiment that optimizes accuracy, we test 1-day, 1-week and
1-month respectively and run the trading method only with sentiment related
to the 500 stocks in the index. The results are shown in
Figure~\ref{fig:tw1}-\ref{fig:tw2}.  For SeekingAlpha, acting on weekly
sentiment is clearly the sweet spot. Acting on daily sentiment over-reacts,
while acting on monthly sentiment is too slow to respond to market changes.
For StockTwits, different $K$ values have minimal impact, but daily
outperforms weekly by a narrow margin.  We use weekly sentiment aggregation
in our later trading strategies.

In summary, our analysis shows that sentiment to performance correlation is
quite low for both SeekingAlpha and StockTwits.  However, there are authors
who consistently provide high-correlation analysis in their articles.  The
challenge is to identify them efficiently.  In the next section, we address
this challenge and develop practical sentiment-driven strategies for stock
trading that can significantly outperform the market.

\begin{figure*}[t]
\centering
% \begin{minipage}{0.64\textwidth}
 \subfigure[SeekingAlpha (2006-2014)]{
  \label{fig:invest1}
  \includegraphics[width=0.4\textwidth]{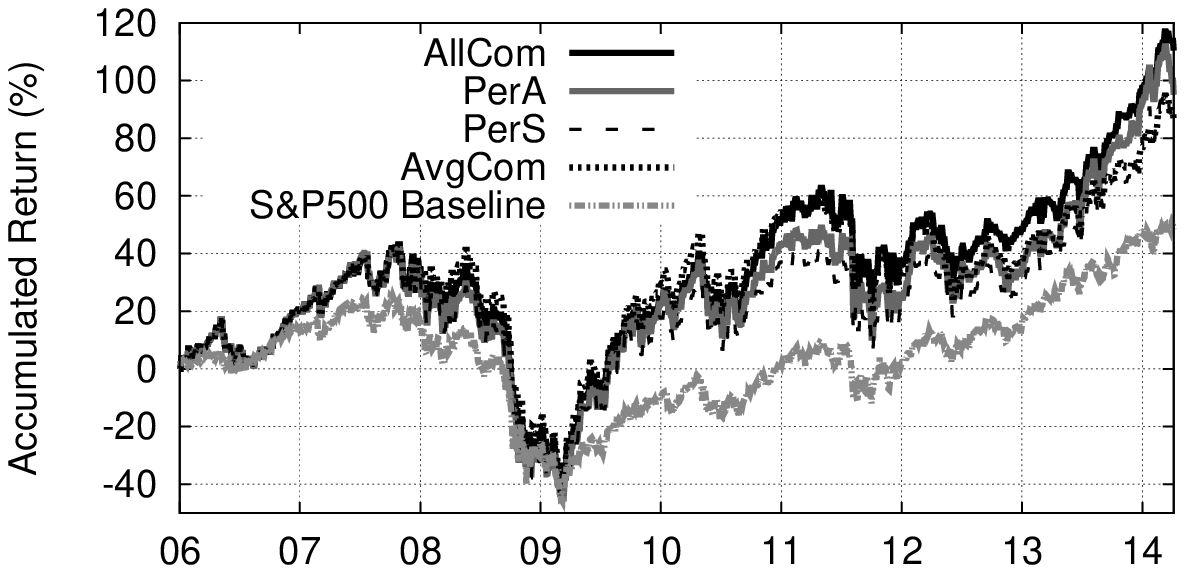}}
\vspace{-0.01in}
  \subfigure[StockTwits (2010-2014)]{
  \label{fig:invest2}
  \includegraphics[width=0.4\textwidth]{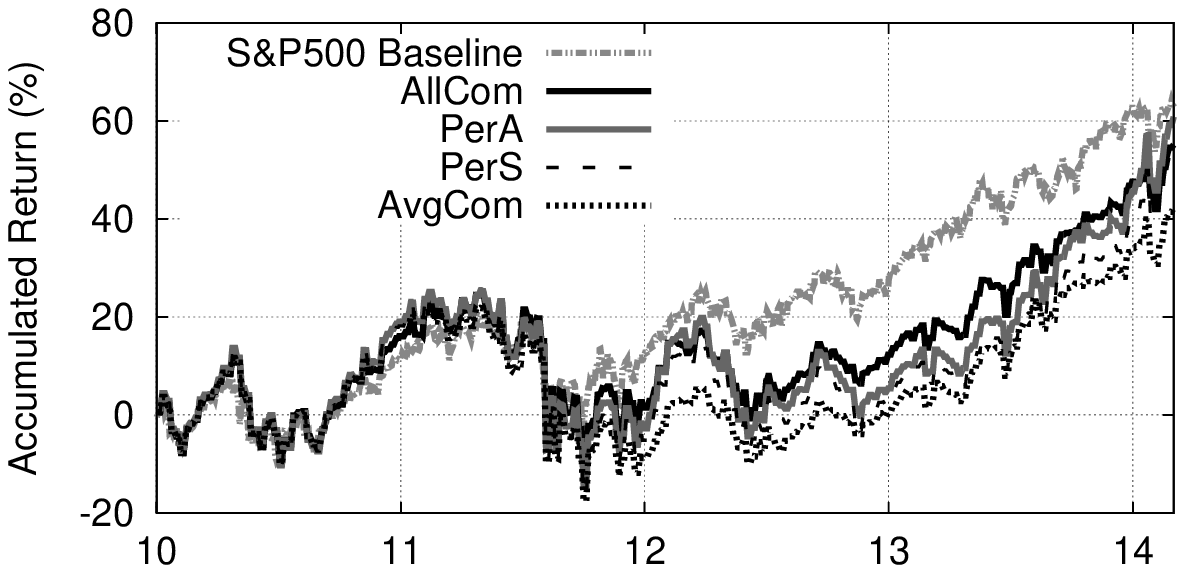}}
\vspace{-0.01in}
\caption{Investing results on 500 stocks recommended by top authors
  over the years.}
\label{fig:invest}
% \end{minipage}
\vspace{-0.03in}
\end{figure*}

% More specifically, we leverage the data from SeekingAlpha
% and StockTwits to assist two key investment tasks: selecting promising
% stocks, and guiding the trading actions. By evaluating the strategy
% return under different scenarios, we week to

\section{6. Practical Sentiment-based Trading}
\label{sec:eval}

Thus far, we have determined that while correlation to stock performance is
low over all articles as a whole, certain subsets of the user population
contribute content with much stronger correlation to stock performance.  Two
key questions remain. First, can these ``valuable'' authors be identified
easily?  Second, once identified, can their content be analyzed to form the
basis of real stock trading strategies, and how would such strategies perform?

We address these questions in three parts.  First, we explore several possible
ranking heuristics for identifying the valuable authors (and their analysis
contributions) from both SeekingAlpha and StockTwits.  Second, we consider
possible stock trading strategies based on sentiment analysis of these
contributions.  Finally, we use historical stock data to drive simulations of
these trading strategies, and use empirical results to draw conclusions about
the value of these top authors, and the efficacy of our mechanisms to
identify them.

\subsection{6.1 Ranking Authors}
% A key lesson from previous section is that we need to identify
% reliable authors for useful opinions and predictive sentiment. Thus
% ranking authors becomes a key step in our investment strategy. 
To identify the (possibly small) subset of top authors in our systems, we
explore two different sets of heuristics.  First, we consider using empirical past
performance, {\em i.e.} correlation between sentiment and stock performance,
as a gauge to rank authors.  While this is likely the most direct way to rank
authors by performance, its computation requires access to significant
resources, including past stock data and sentiment analysis tools.  Second,
we consider a simpler alternative based on user interactions (comments).  The
intuition is that user feedback and engagement with content provides a good
indicator of valuable content.

% empirically evaluating authors' previous prediction accuracy, or using
% user interactions to identify socially popular authors.
% Both approaches rely on authors' (recent) historical data to perform
% ranking. Once we have ranked authors, we can follow top authors to
% proceed investment, more specifically, selecting stocks mentioned by
% top authors and trading stocks based on top author sentiments.

\para{Ranking Authors by Prediction Accuracy.} Our first ranking heuristic is
purely empirical: we rank authors based on how well their previous articles
predict stock returns. For a given author and historical time period ({\em
  e.g.} a year), we compute average hypothetical return of her articles
posted during that given period. Recall that we used this in Section~5 as a
metric of an author's prediction ability.  A variant of this ranking metric
is an author's average hypothetical return {\em per stock}. Compared to the
{\em per article} metric, this highlights authors who have consistently good
performance over a range of stocks over those who write numerous articles on
a small set of stocks. We consider both metrics in our experiments.
 
% For each article, we measure its prediction correctness based
% on the hypothetical return: If the article is positive about the stock, the return  $R = \log
% (\frac{P(d+W)}{P(d)})$, where $P(d)$ is the closing price
% of the day when the article was posted and $W$ is the delay of days to
% assess the correlation.  Similarly, if the sentiment is negative, $R =
% -\log (\frac{P(d+w)}{P(d)})$.  Then the author's expertise score is
% measured by the average return per article. The key parameter is the
% delay $W$, which we set as 1-week (see last Section).

\para{Ranking Authors by Received Comments.} The challenge with empirical
performance-based metrics is that it requires significant resources in
historical data and computation.  Here, we also consider the value of a
simpler approximation based on reader engagement.  The intuition is that the
audience in these systems is a valuable asset, and we can observe reader
responses to contributed content and indirectly infer the value of the
content.  More specifically, we use two heuristics that rank authors
based on either total number of comments or comments per-article.  Without semantic
analysis of comments, we use the number of comments as an approximate
indicator of user agreement.

\subsection{6.2 Sentiment-based Stock Trading Strategies}
Given a ranking of top authors, the next step is to formulate a stock trading
strategy that takes advantage of (hopefully) valuable and predictive
sentiment on individual stocks. Our strategies build up on these articles by
selecting stocks mentioned by top authors in their (recent) articles.  For
simplicity, we build a portfolio for our simulations from the 500 stocks
mentioned by the top-ranked authors.  Experiments with smaller portfolios
show highly consistent results, and are omitted for brevity.

In terms of trading strategies, we implement two simple strategies: 
a basic ``long'' strategy (buy or sell based on sentiment) similar to the one
used to trade the {\tt SPY} in Section~5.2, and a more aggressive
``long/short'' strategy that allows investors to short stocks. For both
strategies, we trade stocks on a weekly basis based on earlier results
(Figures~\ref{fig:tw1}--\ref{fig:tw2}). 

% Here we use author's (negative) sentiment to help investors decide
% when to give up the stock. 
\para{Long Strategy.} Our long strategy builds a portfolio by initially
spreading funds evenly to purchase $N$ stocks, $N=500$ for our examples.
Then we make trading decisions on each stock independently on a weekly
basis. For a stock in the portfolio, we sell our entire position in the stock if the
aggregated sentiment about this stock in the past week is
negative. Otherwise, we hold the stock (or buy it back if we sold it earlier).
We use the same sentiment aggregation method as before (\S 5.2),
but only consider the top author's sentiment on each stock. The return
of the portfolio is the sum of returns over all stocks.

\para{Long/Short Strategy.}  A more aggressive ``long/short'' strategy not
only buys stocks with positive sentiment, but also proactively
``shorts'' stocks with negative sentiments.  Investors ``short'' a stock they
believe will drop in value, by ``borrowing'' stock shares and selling at the
current price, then buying the stock shares back later at a lower price.  The
investor earns the price difference after a price drop.  Shorting is
generally considered to be very risky, because the price of a stock can go up
without limit, thus there is no limit to the size of potential losses on a short
position.

In our short strategy, if the aggregated sentiment on a stock is negative in
previous week, we not only sell any shares of the stock in our portfolio, but
we also short the stock for a week (and buy shares back at the end of the
week). We short a number of shares equal to value to $\frac{1}{N}$ of our
total portfolio.  If and when we have lost 100\% of the value initially
allocated to a stock, then we close our position on that stock and remove it
from our portfolio.

\subsection{6.3 Empirical Evaluation}
We evaluate trading strategies generated using a combination of author
ranking heuristics and long vs. long/short trading strategies.  For our
author ranking heuristics, we use average return per article ({\em PerA}),
average return per stock ({\em PerS}), number of total comments ({\em
  AllCom}), and average comments per article ({\em AvgCom}). As described
above, we choose 500 stocks mentioned by the top ranked authors, and split
the funds of a hypothetical portfolio evenly among them.  Each week, we trade
them based on aggregated sentiment from our chosen top authors from the
previous week. By default, we regard one year as a cycle, and rerank authors
and reset the list of stocks for the portfolio at the beginning of
each year (using the heuristics of previous year).

{\em Q1: Do our strategies outperform the broader markets?}
 
We simulate our strategies using historical stock price data and compare them
to a baseline following a ``buy-and-hold'' strategy on the S\&P 500 index.
We ignore transaction fees in our simulations and initially focus on the
long-only strategy.  We plot results in Figure~\ref{fig:invest}.

The first takeaway is that SeekingAlpha clearly out-performs the baseline
market under all settings. For example, our ``all comment'' strategy produces
a normalized total return of 108\% at the end of the 8-year period, compared
to 47.8\% of the S\&P~500.  This represents more than 10\% annual compounded
return, during a time period that includes two market crashes (2008, 2011),
and {\em not including dividends}.  Since StockTwits only started in 2009, its
simulations ran on only 4 years of historical data.  The same strategy on
StockTwits produced a total return of around 54.5\% from 2010 to 2014.  This is
a good return in absolute terms, but significantly below the baseline
S\&P~500 (64.1\%) during the same timeframe. 

{\bf Implications.} This result is {\em significant}, because it means we can in fact use 
empirical methods to identify the articles of value from SeekingAlpha.  More
importantly, we {\em significantly} outperform the S\&P~500 using a very
simple trading strategy that ignores semantic meaning of the articles and
uses only binary sentiment values.  As context, we consider hedge funds,
which manage money for large investors, and charge annually 2\% of assets
managed and 20\% commission on all gains. The 2.5 {\em trillion} dollar hedge
fund industry has underperformed the S\&P~500 for 5 years in a row, and has
fallen behind the S\&P by 97\% since 2008~\cite{hedge}.  Significantly
outperforming the S\&P over a period of 10 years would be considered
excellent performance for a managed hedge fund.  The fact that this is
achieved by mining an open community like SeekingAlpha is quite surprising.

{\em Q2: What ranking method identifies experts most effectively?} 

The next big question is can we validate a simple methodology for identifying
the top performing authors.  Among a number of author ranking heuristics, we
find the all-comment metric ({\em R-allC}) to obtain the highest level of
investment returns in our SeekingAlpha simulations (see
Figure~\ref{fig:invest}).  Similarly, the same strategy also performs the
best for StockTwits, outperforming other metrics in most years. 

{\bf Implications.} This result implies that not only does something as
simple as comment count do a great job of identifying top authors in
SeekingAlpha, but it does even better than heuristics based on prior-year
performance.  Note that we are not leveraging sentiment analysis on comments,
only the number of comments.  This implies that the majority of comments are
supportive notes that validate the article's analysis and insights.  More
importantly, this highlights the value of the SeekingAlpha user population as
a filter to identify the best analysts among the authors.  Even for a
subject as complex and domain specific as stock analysis, a reasonably
knowledgeable crowd can serve to pinpoint valuable content in an otherwise
mixed collection of content.
% itself is controversial. These stocks that draw heated discussion
% are more likely to be promising investment opportunities. 
% In addition, we find the per-article return metric ({\em R-A})
% achieves better performance than per-stock metric ({\em R-S}). 
% This indicates authors typically have consistent prediction
% performance across the different stocks, thus {\em R-S} does not
% offer the benefit we expected.

\begin{figure}[t]
\centering
  \subfigure[SeekingAlpha (2006--2014)]{
  \label{fig:short1}
  \includegraphics[width=0.23\textwidth]{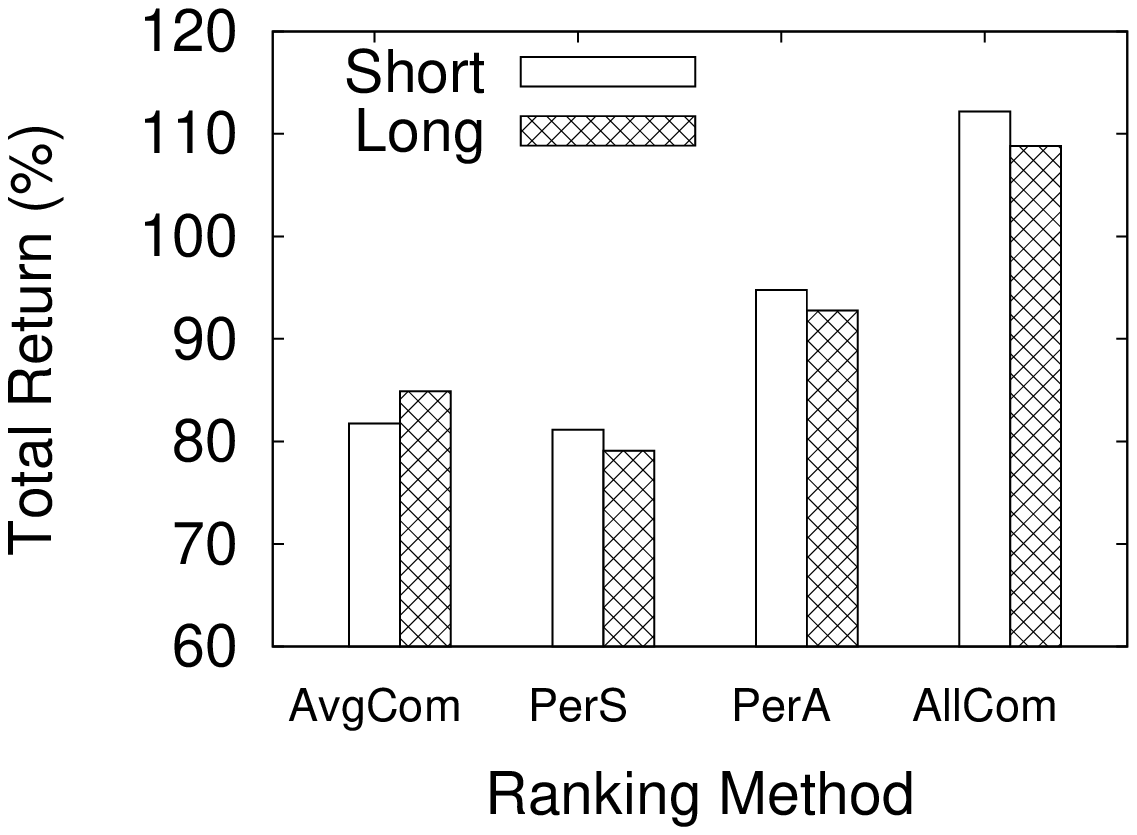}}
  \subfigure[StockTwits (2010--2014)]{
  \label{fig:short2}
  \includegraphics[width=0.23\textwidth]{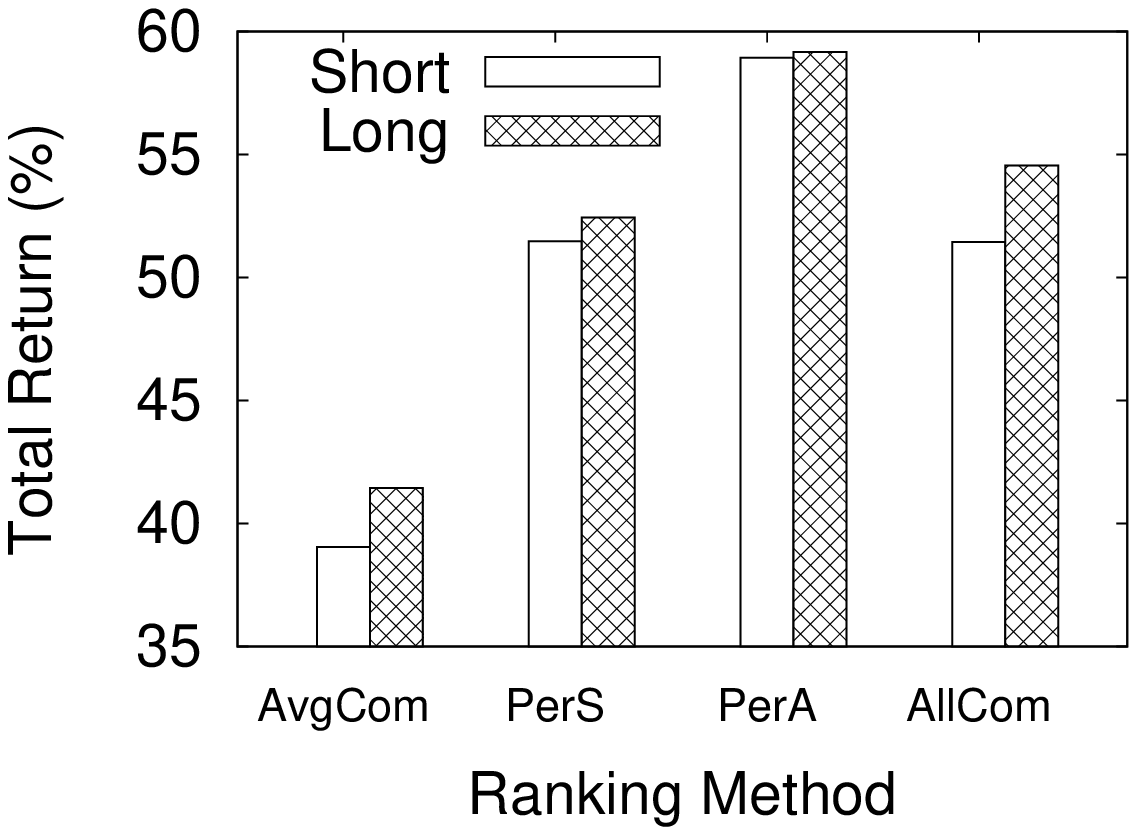}}
  \caption{The total return with long strategy versus short strategy.}
 \label{fig:short}
\vspace{-0.03in}
\end{figure}

{\em Q3: Do more aggressive strategies improve performance? }

We also study the impact of allowing shorts of stocks along with traditional
long positions.  Shorting stocks is a typical strategy used by hedge funds,
but rarely used by individual investors because of its potential for
unbounded loss.  We repeat the same experiment as above, but add a long/short
strategy in addition to the long strategy. The results are plotted in
Figure~\ref{fig:short} (note the different time frames in the two
subfigures).  The high level observation is that adding shorts does improve
performance for trading strategies based on SeekingAlpha, the improvement is
not significant enough to justify the added risk.

{\bf Implications.}  Shorting stocks is a highly valued tool for hedge funds,
who are expected to use it to produce positive returns even in negative
markets. Yet our results show that the majority of user-contributed short
strategies do not produce significant gains over long-only strategies.  Given
the significant added risk, this suggests that sentiment-based trading
strategies should focus on long-only strategies to minimize risk while
achieving the gains of a long/short strategy.

In summary, our measurement shows by carefully identifying top authors, data
from SeekingAlpha can provide significant investment returns that
consistently outpace the broader market.  We also find that through their
interactions and article comments, the broader user population can provide an
effective filter for top authors and content.

\section{7. SeekingAlpha User Survey}
\label{sec:user}
The final component of our study deploys a user survey on SeekingAlpha to
better understand users' levels of investment experience, and how they feel
about the utility and reliability of SeekingAlpha articles.  These responses
will help us understand the level of impact articles have on SA users, and
how they deal with any potentially manipulative articles.

% of the benefits as well as concerns in using the platform. More specifically,
% we conduct user survey on both authors and non-authors to answer two key
% questions: First, do users recognize the value of SeekingAlpha articles, and
% do they actually follow the investing ideas in the articles for their own
% investment? Second, given the influence of platform, do people have concerns
% on potential stock manipulation behaviors ({\em e.g.}, heavily biased
% articles with the intent to impact stock price)?

\subsection{7.1 Survey Setup}

In May 2014, we sent out a user survey via private messages in SeekingAlpha
to 500 authors and 500 non-contributing users. We received 199
responses (95 from authors and 104 from normal users).  
We chose authors and users to ensure we captured a full range
of activity levels. We sorted authors into buckets by the number of articles
written and users into buckets by the number of comments, and randomly sampled
from each bucket to choose our targets.

We asked normal users 5 questions and authors 4 questions (see the full
questions in Table~\ref{tab:qlist} in the Appendix).  First, to measure
demographics, we ask both authors and users about their levels of investing
experience (Q1). Second, to understand user's perceived value of the
platform, we ask whether normal users trade stocks based on SA articles (Q2)
and whether they trust authors (Q3). For authors, we ask which platforms if
any they would use to disseminate investing ideas if SeekingAlpha were no
longer available (Q7). Third, in terms of risks in the platform, we ask both
authors and users whether they have seen heavily biased (potentially
manipulative) articles (Q4).  Finally, we ask normal users for their reaction to
stock manipulation (Q5), and authors whether they believe articles can actually
impact stock price (Q6).

\begin{figure}
\centering
  \subfigure[Years of investing experience]{
  \label{fig:qcom1}
  \includegraphics[width=2in]{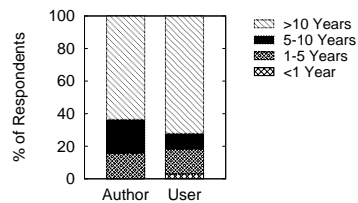}}
  \subfigure[Seen manipulation articles?]{
  \label{fig:qcom2}
  \includegraphics[width=2in]{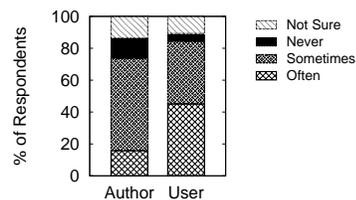}}
  \caption{Results of two questions for both authors and users.}
 \label{fig:qcom}
\vspace{-0.03in}
\end{figure}

\subsection{7.2 User Responses}
We first look at demographics of the survey respondents. We plot responses to
Q1 in Figure~\ref{fig:qcom1}. The large majority of respondents have over 10
years of investing experience. Surprisingly, we find that the portion of
experienced investors ({\em i.e.} $>$10 years) is higher in normal users than
in authors.

\para{Perceived Value.} First, in response to Q2, 70\% of users stated that
they ``sometimes'' or ``very often'' traded stocks following opinions from
SeekingAlpha. 5\% of users stated they ``always'' relied on information from
SA articles.  The remaining 25\% ``never'' followed views from articles. Two
respondents added clarifying comments that they treated SeekingAlpha as an
important information source for research, but did not follow opinions
blindly. As a whole, it is clear that SeekingAlpha articles play an important
role in most users' investment decisions.

In response to Q3, 83\% of users stated they trusted ``some'' or ``most''
SeekingAlpha authors, and 1\% stated they fully trusted
authors. 16\% did not trust authors at all. This suggests that while
SeekingAlpha users generally acknowledge the value of the platform,
hesitations and concerns remain.

Q7 asked authors where they would disseminate their ideas if
SeekingAlpha were no longer available.  Surprisingly, only 6\% of authors chose StockTwits,
and more chose Twitter (7\%), Yahoo! Finance Message Board (9\%) and Motley
Fool (22\%).  Still, the majority of authors (56\%) did not choose any
existing platform, but preferred a personal website or nothing at all.
Clearly, SeekingAlpha provides a unique and valuable platform for its
contributors.

\para{Biased Articles and Stock Manipulation.} SeekingAlpha's nature makes it
possible for companies or individuals to profit by influencing users and
manipulating stock prices. Some of these efforts were identified and publicly
documented~\cite{alpha4,alpha3}.  Surprisingly, both authors and users (80\%)
stated that they have seen manipulative articles on SA
(Figure~\ref{fig:qcom2}).  More normal users reported frequently seeing
manipulative articles.  When asked about their response to manipulative
articles (Q5), 42\% stated that they would dispute the conclusions using
comments on the article, 3\% would report to SeekingAlpha admins, while 25\% ``do
nothing.''  20 users added further detailed comments to explain: 15
respondents stated biased articles are to be expected and ignored; 2 people
said they would blacklist the author and never read their articles again; 2
stated they could still extract value from dissenting comments on the
article. Finally, one user stated that it was very difficult to distinguish
between real stock manipulation articles and those from authors with strong
personal preferences.

\para{Impact.} We ask if authors believe their articles have the
power to impact stock prices (Q6). More than 62\% of authors believed their
articles could impact stock prices; 14\% said no, and the rest were
unsure. One author commented that SeekingAlpha articles typically cannot
impact large cap stocks like Apple ({\tt AAPL}), but they could affect small
and micro cap stocks.

Finally, we also received unsolicited anecdotal feedback from past authors on
additional types of author misbehavior.  They identified some authors who
generated large volumes of articles solely for the purpose of soliciting
users to subscribe to their investment website memberships. They also identified
others who produced detailed articles at extremely high volumes, too
high to be produced even by a full time professional.  The assertion is these
accounts are managed by teams of writers who generate numerous articles to
increase in reputation and ranking, only to then profit by manipulating one
or two stocks they own in large volume. While we cannot confirm or disprove
these assertions, they are consistent with other survey responses, and could
account for the apparent disconnect between poor average correlation of
articles and high correlation of articles by top authors (Section~6).

\section{8. Related Work}
\label{sec:relate}
\para{Stock Market Prediction.}
% Related work using Twitter, Google trend, Yahoo Finance news to
% predict the future trend of stock market. 
% Related work using StockTwits and seeking Alpha data to predict market
% trend and earning surprise 
%larger data set; measurement, understand the insights
%practical investment strategy
Stock market prediction via data mining has been explored in a variety of
contexts, including Google trends~\cite{preis2013quantifying},
Wikipedia~\cite{moat2013quantifying}, online
blogs~\cite{de2008can,gilbert2010widespread}, financial
news~\cite{fung2003stock,schumaker2009textual}, and social content from
Twitter~\cite{bollen2011twitter,brown2012will,rao2012analyzing,
  sprenger2013tweets} and Facebook~\cite{karabulut2011can}.  Some have
studied stock-centered social networks, {\em i.e.}
StockTwits~\cite{oh2011investigating, oliveira2013predictability} and
SeekingAlpha~\cite{chen2013wisdom}.  Most of these draw their conclusions
based on short-term data of less than a year, despite the highly cyclical
nature of bull and bear markets that lasts multiple years.  In contrast, our
data covers up to 9 years, long enough to cover both crashes (2008--2009) as
well as strong bull markets (2013).  In addition, our comparative analysis
between StockTwits and SeekingAlpha helps us understand the impact of
leveraging experts as contributors versus average users.
%authors could impact the quality of the prediction.

\para{Sentiment-based Investment Strategies.}
There are a few works on sentiment based investment strategy. Most
papers focus on the prediction and investment on several selected
stocks~\cite{oliveira2013predictability}. Other
work~\cite{makrehchi2013stock} invests on all possible stocks in the
market. Prior works~\cite{bar2011identifying,liao2014winning} use small
datasets from Twitter/StockTwits to quantify the high level of noise
in overall sentiment, and to motivate the need for contributions from
experts. This is consistent with our results that show the best performance is
achieved from identifying and relying on top experts.

\para{Sentiment Analysis.}
Existing sentiment analysis methods vary widely from dictionary based
methods~\cite{esuli2007pageranking,godbole2007large,loughran2011liability} to
supervised machine learning
algorithms~\cite{liu2012sentiment,pang2002thumbs,barbosa2010robust}.
Dictionary based methods require domain specific
dictionaries~\cite{loughran2011liability} instead of general dictionaries
such as WordNet~\cite{esuli2007pageranking}, and do not work well with short
texts like tweets~\cite{barbosa2010robust} and
reviews~\cite{pang2002thumbs}. Researchers have applied supervised machine
learning algorithms to sentiment classification, with common features such as
term frequency, parts of speech, and negations~\cite{liu2012sentiment}.

\para{Crowd-based Stock Prediction.} Finally,
Estimize\footnote{\url{http://www.estimize.com}} is an open financial platform
that aggregates estimates of company earnings reports from the opinions of
independent, buy-side, and sell-side analysts, along with private investors.
Estimize has contributions from 4628 analysts covering over 900 stocks.  Note
that Estimize focuses on predictions of quarterly earnings results, not stock
performance.

\section{9. Conclusions and Future Work}
\label{sec:con}

In this paper, we analyze the correlation between stock performance and user
contributed content sentiment over a period of 4--9 years.  Our analysis
shows that while expert-contributed stock analysis in SeekingAlpha provides more
positive correlation than user-generated content from StockTwits, the correlation is
very weak.  We show that valuable content can be extracted using well
designed filters based on user comments, and can lead to strategies that
significantly outperform the broader stock market.  Future work lies in
better identification of biased or manipulative content on SeekingAlpha, as
well as further analysis of how to extend such methods to other platforms.

Our work suggests that even complex, domain specific tasks such as stock
analysis can benefit from crowd-contributed content.  While only a small
portion of users can deliver valuable content, others in the crowd help by
indirectly identifying the valuable content through their interactions.  This
bodes well for the application of crowdsourcing platforms to broader expert
applications in the future.

% \newpage
\begin{small}
\balance
\bibliographystyle{acm-sigchi}
\bibliography{zhao,astro,seekingalpha}
\end{small}

\section*{Appendix A -- SeekingAlpha Survey}

These are the questions sent out in our questionnaire to SeekingAlpha users
and contributors.  We focused on limiting the number of questions to improve
our chances of getting responses.  Authors or contributors were asked four
questions (\#1, \#4, \#6, and \#7 in Table~\ref{tab:qlist}), while
non-contributing users were asked five questions (\#1, \#2, \#3, \#4, and
\#5). 

\begin{table}[h]
\small{
\centering{
\begin{tabular}{|p{0.15cm}|p{6.5cm}|p{0.9cm}|}
\hline
 	 ID     &
	 Question &
	Tester \\
\hline
	1	&
        How many years of experience do you have in investing in the stock market?   & 	
	Author \& User \\
\hline
	2	&
       How often do you trade stocks based on opinions and information
       gathered from reading SeekingAlpha articles? & 	
	User \\
\hline
	3	&
      Do you think you can trust the authors on SeekingAlpha?  & 	
	User \\
\hline
	4	&
        How often have you seen articles on SeekingAlpha that looked
        like they were heavily biased, written with the intent to
        manipulate a particular stock (to move its price up or down)?  & 	
	Author \&  User \\
\hline
	5	&
       If and when you did see what looked like an article intent on
       stock manipulation, what is your general reaction?   & 	
       User \\
\hline
	6	&
       Do you think SeekingAlpha articles can impact the future
       movement of stocks they focus on?  & 	
	Author \\
\hline
	7	&
       If SeekingAlpha did not exist, where would you post your
       investment ideas?  & 	
	Author \\
\hline
\end{tabular}
\caption{Survey questions to SeekingAlpha authors and normal users. }
\label{tab:qlist}
}}
\end{table}

\end{document}